\documentstyle[12pt,psfig]{article}
\catcode`\@=11
\def\gappr{\mathpalette\under@rel{>\approx}}
\def\lappr{\mathpalette\under@rel{<\approx}}
\def\gsim{\mathpalette\under@rel{>\sim}}
\def\lsim{\mathpalette\under@rel{<\sim}}
\def\under@rel#1#2{\under@@rel#1#2}
 
\def\under@@rel#1#2#3{\mathrel{\mathop{#1#2}\limits_{#1#3}}}
 
\def\under@@rel#1#2#3{\mathrel{\vcenter{\hbox{$%
  \lower3.8pt\hbox{$#1#2$}\atop{\raise1.8pt\hbox{$#1#3$}}%
  $}}}}

\def\parenbar{\mathpalette\p@renb@r}
\def\p@renb@r#1#2{\vbox{%
  \ifx#1\scriptscriptstyle \dimen@.7em\dimen@ii.2em\else
  \ifx#1\scriptstyle \dimen@.8em\dimen@ii.25em\else
  \dimen@1em\dimen@ii.4em\fi\fi \offinterlineskip
  \ialign{\hfill##\hfill\cr
    \vbox{\hrule width\dimen@ii}\cr
    \noalign{\vskip-.3ex}%
    \hbox to\dimen@{$\mathchar300\hfil\mathchar301$}\cr
    \noalign{\vskip-.3ex}%
    $#1#2$\cr}}}
\catcode`\@=12 
\newbox\struttbox
\setbox\struttbox=\hbox{\vrule height1.65ex depth.485ex width0pt}
\def\strutt{\relax\ifmmode\copy\struttbox\else\unhcopy\struttbox\fi}
\def\stru#1#2{\relax\ifmmode\hbox{\vrule height#1 depth#2 width0pt}
\else\vrule height#1 depth#2 width0pt\fi}
\def\uline#1{$\underline{\hbox{#1\strutt}}$}
\mathchardef\smallleft=300
\mathchardef\smallright=301

\def\ronum#1{\uppercase\expandafter{\romannumeral#1}}
\def\ronuml#1{\expandafter{\romannumeral#1}}

\newfont{\bm}{cmmib10 scaled 1200}
\mathchardef\pls=43
\mathchardef\mns=512
\mathchardef\plm=518
\mathchardef\eql=61

\def\cha{{\phantom{0}}}
\def\chaa{{\phantom{00}}}

\def\ev{{\rm e}\kern-1.pt{\rm V}}
\def\kev{{\rm ke}\kern-1.pt{\rm V}}
\def\mev{{\rm Me}\kern-1.pt{\rm V}}
\def\gev{{\rm Ge}\kern-1.pt{\rm V}}
\def\tev{{\rm Te}\kern-1.pt{\rm V}}
\mathchardef\less=316
\mathchardef\greater=318
\def\3{\ss}                                                                                        
\def\annu{{\overline\nu}}
\def\ubar{{\overline u}}
\def\dbar{{\overline d}}
\def\sbar{{\overline s}}
\def\cbar{{\overline c}}

\def\chax{{\phantom{0}}}
\def\chapx{{\phantom{.0}}}
\def\chaxx{{\phantom{00}}}

\def\chasx{{\phantom{\mns0}}}
\def\chas{{\phantom{\mns}}}
\def\reaep{{e^+p\to\annu X}}
\def\reaem{{e^-p\to\nu X}}
\def\sigep{\sigma_{e^+p\to\annu X}}
\def\sigem{\sigma_{e^-p\to\nu X}}
\def\sigepnc{\sigma_{e^+p\to e^+ X}}
\def\sigemnc{\sigma_{e^-p\to e^- X}}
\def\sigrat#1{{(d\sigep/d\,#1)/(d\sigem/d\,#1)}}

\def\none{\hbox{---}}
\def\zvtx{{Z_{\rm vtx}}}
\newcommand{\ptmiss}{{\mbox{$\not\hspace{-.55ex}{P}_t$}}}
\newlength{\dinmargin}
\newlength{\dinwidth}
\begin{document}
\vspace{1 cm}
\begin{titlepage}

\title{\bf Study of Charged--Current $ep$ Interactions\\
           at $Q^2>200\,\gev^2$ with the ZEUS Detector at HERA}
\author{ ZEUS Collaboration}
\date{}

\maketitle

\vspace{5 cm}

\begin{abstract}
\centerline{\vbox{\hsize13.7cm
\noindent
Deep inelastic charged--current reactions have been studied in $e^+p$ and $e^-p$
collisions at a center of mass energy of about $300\,\gev$ in the kinematic
region $Q^2\greater200\,\gev^2$ and $x\greater0.006$ using the ZEUS detector at
HERA.  The integrated cross sections for $Q^2\greater200\,\gev^2$ are found to
be $\sigep=30.3\,{}^{+5.5}_{\mns4.2}\,{}^{+1.6}_{\mns2.6}\,{\rm pb}$ and
$\sigem=54.7\,{}^{+15.9}_{\mns\chax 9.8}\,{}^{+2.8}_{\mns3.4}\,{\rm pb}$.
Differential cross sections have been measured as functions of the variables
$x$, $y$ and $Q^2$.  From the measured differential cross sections
$d\sigma/dQ^2$, the $W$ boson mass is determined to be $M_W=79\,{}^{+8}
_{-7}{}^{+4}_{-4}\,\gev$. Measured jet rates and transverse energy profiles
agree with model predictions.  A search for charged--current interactions with a
large rapidity gap yielded one candidate event, corresponding to a cross section
of $\sigep(Q^2\greater200\,\gev^2;\eta_{\rm max}<2.5)=0.8\,{}_{-0.7}^
{+1.8}\,\pm0.1\,{\rm pb}$.
  }}

\end{abstract}

\setcounter{page}{0}                                                            
\thispagestyle{empty}   

\end{titlepage}
\newcommand{\address}{ }                                                                           
\renewcommand{\author}{ }                                                                          
\pagenumbering{Roman}                                                                              
{
\footnotesize
\parskip0.3cm plus0.05cm minus0.05cm

\begin{center}                                                                                     
{                      \Large  The ZEUS Collaboration              }                               
\end{center}                                                                                       
  M.~Derrick,                                                                                      
  D.~Krakauer,                                                                                     
  S.~Magill,                                                                                       
  D.~Mikunas,                                                                                      
  B.~Musgrave,                                                                                     
  J.R.~Okrasinski,                                                                                 
  J.~Repond,                                                                                       
  R.~Stanek,                                                                                       
  R.L.~Talaga,                                                                                     
  H.~Zhang  \\                                                                                     
 {\it Argonne National Laboratory, Argonne, IL, USA}~$^{p}$
% ...
\newpage                                                                                           
% ...
$^{  20}$ now at ATLAS Collaboration, Univ. of Munich \\                                           
% ...
                                                           %
                                                           %
\newpage   
                                                           %
                                                           %
\begin{tabular}[h]{rp{14cm}}                                                                       
$^{a}$ &  supported by the Natural Sciences and Engineering Research                               
          Council of Canada (NSERC)  \\                                                            
$^{b}$ &  supported by the FCAR of Qu\'ebec, Canada  \\                                            
$^{c}$ &  supported by the German Federal Ministry for Education and                               
          Science, Research and Technology (BMBF), under contract                                  
          numbers 057BN19P, 057FR19P, 057HH19P, 057HH29P, 057SI75I \\                              
$^{d}$ &  supported by the MINERVA Gesellschaft f\"ur Forschung GmbH,                              
          the Israel Academy of Science and the U.S.-Israel Binational                             
          Science Foundation \\                                                                    
$^{e}$ &  supported by the German Israeli Foundation, and                                          
          by the Israel Academy of Science  \\                                                     
$^{f}$ &  supported by the Italian National Institute for Nuclear Physics                          
          (INFN) \\                                                                                
$^{g}$ &  supported by the Japanese Ministry of Education, Science and                             
          Culture (the Monbusho) and its grants for Scientific Research \\                         
$^{h}$ &  supported by the Korean Ministry of Education and Korea Science                          
          and Engineering Foundation  \\                                                           
$^{i}$ &  supported by the Netherlands Foundation for Research on                                  
          Matter (FOM) \\                                                                          
$^{j}$ &  supported by the Polish State Committee for Scientific                                   
          Research, grants No.~115/E-343/SPUB/P03/109/95, 2P03B 244                                
          08p02, p03, p04 and p05, and the Foundation for Polish-German                            
          Collaboration (proj. No. 506/92) \\                                                      
$^{k}$ &  supported by the Polish State Committee for Scientific                                   
          Research (grant No. 2 P03B 083 08) and Foundation for                                    
          Polish-German Collaboration  \\                                                          
$^{l}$ &  partially supported by the German Federal Ministry for                                   
          Education and Science, Research and Technology (BMBF)  \\                                
$^{m}$ &  supported by the German Federal Ministry for Education and                               
          Science, Research and Technology (BMBF), and the Fund of                                 
          Fundamental Research of Russian Ministry of Science and                                  
          Education and by INTAS-Grant No. 93-63 \\                                                
$^{n}$ &  supported by the Spanish Ministry of Education                                           
          and Science through funds provided by CICYT \\                                           
$^{o}$ &  supported by the Particle Physics and                                                    
          Astronomy Research Council \\                                                            
$^{p}$ &  supported by the US Department of Energy \\                                              
$^{q}$ &  supported by the US National Science Foundation \\                                       
\end{tabular}                                                                                      
}                                                           %

\newpage

\setcounter{page}{1}
\renewcommand{\thepage}{\arabic{page}}

\setlength{\dinwidth}{21.0cm}
\textheight24.3cm
\textwidth16.0cm
\setlength{\dinmargin}{\dinwidth}
\addtolength{\dinmargin}{-\textwidth}
\setlength{\dinmargin}{0.5\dinmargin}
\oddsidemargin -1.0in
\addtolength{\oddsidemargin}{\dinmargin}
\setlength{\evensidemargin}{\oddsidemargin}
\setlength{\marginparwidth}{0.9\dinmargin}
\marginparsep 8pt
\marginparpush 5pt
\topmargin -42pt
\headheight 12pt
\headsep 30pt
\footheight 12pt
\footskip 24pt
\parindent 0.pt
\parskip 1mm plus 1mm minus 1mm

\section{Introduction}

In comparison to fixed target neutrino scattering experiments \cite{FTneut}, the
HERA $ep$ collider extends the kinematic region for studying charged--current
(CC) deep inelastic scattering (DIS) by about two orders of magnitude in $Q^2$.
In addition, HERA allows measurements of CC DIS at lower $x$ ($x<0.01$). Here
$Q^2$ is the negative square of the four--momentum transferred between the
electron\footnote{In the following, ``electron'' is generically used to denote
both electrons and positrons.} and the proton, and $x$ is the Bjorken scaling
variable.

Both the H1 \cite{H1CC93} and ZEUS \cite{ZEUSCC93} collaborations have
previously reported cross section measurements for the process $e^-p\to\nu_eX$,
where $X$ denotes the hadronic final state. These investigations, based on the
data collected in 1993, established that the $Q^2$ dependence of the CC cross
section is consistent with that of the $W$ propagator and that the CC and
neutral--current (NC) cross sections have similar magnitude for $Q^2\gsim M_W^2$
($M_W$ denotes the $W$ mass).  The H1 collaboration has also measured the
integrated cross section \cite{H1CC94} and the differential cross section
$d\sigma/dQ^2$ \cite{H1CC94Q} for $e^+p\to\annu_eX$ with missing transverse
momentum (\ptmiss) above $25\,\gev$.

In 1994 ZEUS has collected $2.93\,{\rm pb}^{-1}$ of $e^+p$ data and $0.27\,{\rm
pb}^{-1}$ of $e^-p$ data in collisions of $820\,\gev$ protons with $27.5\,\gev$
electrons.  In order to study both $e^+$ and $e^-$ induced CC reactions, these
data samples have been combined with the $0.55\,{\rm pb}^{-1}$ of $e^-p$ data
taken in 1993 with an electron beam energy of $26.7\,\gev$ \cite{ZEUSCC93}.

In the quark parton model (including the leading order QCD evolution in $Q^2$),
the unpolarized $e^+p$ and $e^-p$ charged--current differential cross sections
are:\footnote{For the current integrated luminosity, any contributions to the CC
cross section from hard subprocesses involving $3^{\rm rd}$ generation quarks
($b,t$) can be neglected. In this approximation the relation between quark
densitites and CC structure functions is given by $F_2^\reaep\propto
d+s+\ubar+\cbar$, $F_2^\reaem\propto u+c+\dbar+\sbar$, $xF_3^\reaep\propto
d+s-\ubar-\cbar$ and $xF_3^\reaem\propto u+c-\dbar-\sbar$.}

\begin{equation}
\frac{d^2\sigep}{dx dQ^2} = \frac{G_F^2}{2 \pi}
 \left(\frac{M_W^2}{M_W^2 + Q^2}\right)^2\,
 \sum_{i=1}^2\,\left[x\overline{u}_i(x,Q^2) + (1-y)^2\,xd_i(x,Q^2)\right]\;,
\label{Opos}
\end{equation}

\begin{equation}
\frac{d^2\sigem}{dx dQ^2} = \frac{G_F^2}{2 \pi}
 \left(\frac{M_W^2}{M_W^2 + Q^2}\right)^2\,
 \sum_{i=1}^2\,\left[xu_i(x,Q^2)+ (1-y)^2\,x\overline{d}_i(x,Q^2)\right]\;,
\label{Oele}
\end{equation}

\noindent
where $u_i$ and $d_i$ are the densities of up--type and down--type quarks of the
$i^{\rm th}$ generation in the proton, $G_F$ is the Fermi coupling constant, and
$y$ is the fractional energy transfer to the proton in its rest system. The
factor $(1-y)^2$, which suppresses the quark (antiquark) contribution to the
$e^+p$ ($e^-p$) cross section, is a direct consequence of the $V\mns A$
structure of the weak coupling.

As equations \ref{Opos} and \ref{Oele} show, the charged current couples to
different quark flavors for $e^+$ and $e^-$ beams and the valence and sea
contributions have very different $y$ dependences. In addition, the $Q^2$
dependence of the cross section allows a measurement of the $W$ propagator
effect and thus a determination of the $W$ mass in the space--like region,
complementing the measurements of direct $W$ production at $p\overline p$
colliders \cite{mwppbar}.

This paper reports on measurements of the integrated and differential cross
sections for both $e^+p$ and $e^-p$ CC DIS in the kinematic region
$Q^2>200\,\gev^2$.  The differential cross sections were measured as functions
of $x$, $y$ and $Q^2$, where the latter dependence was used to determine the $W$
mass. The available statistics allowed studies of the hadronic final state at
high $Q^2$ in CC DIS: the jet rates and transverse energy jet profiles were
determined and compared to Monte Carlo predictions. The number of events with
two jets (plus the proton remnant) directly measures the rate of hard QCD
subprocesses.

Recently, diffractive NC DIS has received particular attention
\cite{lrgdis}--\cite{ZEUSDiffXS}. These reactions are characterized by an
absence of final state particles in a wide rapidity interval between the
outgoing proton system (which escapes through the forward beam hole) and the
rest of the hadronic final state. Such a rapidity gap results in a quiet region
in the forward part of the main detector.  Diffractive CC reactions are
interesting because they predominantly occur at high $Q^2$ and also because they
provide the possibility to study the flavor content of the states exchanged
between proton and $W$ boson.  A search for events with a rapidity gap in our CC
samples yielded one candidate, which will be described in detail. The rate of
large rapidity gap CC events was compared to model predictions and to the NC
case.

After a brief summary of the experimental setup in section~2, the kinematics and
the selection of CC events are described in section~3. The cross section
calculations and the evaluation of the systematic uncertainties are outlined in
section~4. Finally, section~5 contains the results, which are summarized in
section~6.

\section{Experimental Setup}

In 1994, a total of 153~colliding bunches were stored in the electron and the
proton beams of HERA, together with an additional 17~proton and 15~electron
non--colliding bunches which were used to study beam induced backgrounds.  The
r.m.s.\ bunch length was about $20\,{\rm cm}$ for protons and $1\,{\rm cm}$ for
electrons, resulting in an event vertex position distribution\footnote{The ZEUS
coordinate system is right--handed with the $Z$ axis pointing in the proton
direction, hereafter referred to as forward, and the $X$ axis horizontal,
pointing towards the center of HERA.} with an r.m.s.\ width of $12\,{\rm cm}$ in
$Z$.  The typical instantaneous luminosity was $1.5\cdot 10^{30}\,{\rm
cm}^{-2}\,{\rm s}^{-1}$. The following description refers to the 1994 running
period. Details of the 1993 experimental setup and HERA running conditions can
be found in \cite{ZEUSF293}.

\subsection{The ZEUS Detector}

A description of the ZEUS detector is available in \cite{ZEUS,trigger}. The
primary components used in this analysis were the uranium--scintillator
calorimeter \cite{CAL} and the central tracking detectors.

The calorimeter is divided into three parts, forward (FCAL) covering the polar
angle\footnote{The polar angle $\theta$ is defined here with respect to the
nominal interaction point.} interval $2.6^\circ\less\theta\less 37^\circ$, the
barrel (BCAL, $37^\circ\less\theta\less129^\circ$) and rear (RCAL,
$129^\circ\less\theta\less176.1^\circ$).  The calorimeters are subdivided into
towers which subtend solid angles between $0.006$ and $0.04$ steradians.  Each
tower is longitudinally segmented into electromagnetic (EMC) and two hadronic
(HAC) sections (one in RCAL).  The electromagnetic layer of each tower is further
subdivided transversely into four cells (two in RCAL).  The total number of
cells is 2172, 2592 and 1154 in FCAL, BCAL and RCAL, respectively.  Under test
beam conditions, the calorimeter has an energy resolution of $\sigma_E/E\eql
18\%/\sqrt{E(\gev)}$ for electrons and $\sigma_E/E\eql35\%/\sqrt{E(\gev)}$ for
hadrons. The time resolution is below $1\,{\rm ns}$ for energy deposits greater
than $4.5\,\gev$.

Energy which penetrates through the CAL can be measured in the Backing
Calorimeter (BAC) which consists of proportional chambers interleaved with the
iron plates which form the return yoke of the solenoid \cite{BAC}.\footnote{Due
to the present limited understanding of the Monte Carlo simulation of energy
deposits in the BAC, this component has been used only for systematic checks in
this analysis.}

The tracking system consists of a vertex detector \cite{VXD} and a central
tracking chamber (CTD) \cite{CTD} operating in a $1.43\,{\rm T}$ magnetic field
parallel to the beam axis.  The polar angle region in which the CTD allows
accurate momentum measurement is $15^\circ\less\theta\less 164^\circ$. The
transverse momentum resolution for full length tracks is $\sigma (p_t)/p_t\eql
\left[0.005p_t(\gev)\right]\oplus0.016$.

The luminosity was measured by the rate of high energy photons from the process
$ep\to ep\gamma$ detected in a lead--scintillator calorimeter \cite{LUMI} located
at $Z=-107\,{\rm m}$.

\subsection{Trigger Configuration}

Data were collected with a three--level trigger system \cite{trigger}. The
first--level trigger conditions used in this analysis were based on
electromagnetic energy, transverse energy, and missing transverse energy in the
calorimeter.  The thresholds were significantly below the corresponding cuts
applied in the subsequent selection.  The second level trigger rejected
backgrounds (mostly $p$--gas interactions and cosmic rays) for which the
calorimeter timing was inconsistent with an $ep$ interaction. Events were
accepted as CC candidates if the following conditions were all fulfilled:
\begin{enumerate}
\item
The missing transverse momentum $\ptmiss$ exceeded $9\,\gev$ (see equation
\ref{ptmissdef}).
\item
$\ptmiss$ evaluated from all calorimeter cells except those adjacent to the
forward beam pipe exceeded $8\,\gev$.
\item
Either a track was found in the CTD or at least $10\,\gev$ was deposited in the
FCAL.
\end{enumerate}

\noindent
The third--level trigger applied stricter timing cuts and also pattern
recognition algorithms to reject cosmic rays.

\subsection{Monte Carlo Simulation}

Simulated CC DIS events with electroweak radiative corrections were generated
using {\sc lepto} \cite{lepto} interfaced to {\sc heracles} \cite{heracles} via
{\sc django} \cite{django}. The MRSA set of NLO parton density parameterizations
was used, which is based on preliminary 1993 $F_2$ measurements by ZEUS and H1
as well as on recent data on $W$ asymmetries from CDF and on Drell--Yan cross
sections (cf.~\cite{mrsa} and the references therein).  The latter measurements
constrain the difference $\ubar-\dbar$, to which CC reactions are particularly
sensitive.  The hadronic final state was simulated using the color--dipole model
as implemented in {\sc ariadne} \cite{ariadne} for the QCD cascade and {\sc
jetset} \cite{pythia} for the hadronization. In addition, MC samples using the
{\sc meps} option of {\sc lepto} instead of {\sc ariadne} were generated. Each
of the CC MC samples contained about $10^4$ events. The results were corrected
for radiative effects using these CC MC sets\footnote{All kinematic quantities
are inferred from the four--momentum of the propagator.}.

MC samples of NC DIS and photoproduction events were employed to simulate
backgrounds to CC DIS.  For the NC DIS simulation the same programs were used as
for the CC events.  Samples of both direct and resolved photoproduction events
were generated using both {\sc pythia} \cite{pythia} and {\sc herwig}
\cite{herwig}.  Photoproduction of $c{\overline c}$ and $b{\overline b}$
pairs was simulated using both {\sc pythia} and {\sc aroma} \cite{aroma}.

To calculate efficiencies for diffractive CC interactions the {\sc pompyt}
\cite{pompyt} Monte Carlo program was used. This program is based on the
factorizing model of Ingelman and Schlein \cite{Schlein} and assumes a hard
quark density in the pomeron which satisfies the momentum sum rule.

All simulated events were passed through a {\sc geant} based \cite{geant}
detector simulation and processed with the same analysis programs as the data.

\section{Kinematics and Event Selection}

In CC $ep$ reactions the final state neutrino remains undetected. This fact
imposed special conditions both for the reconstruction of kinematic variables
and for the event selection.

\subsection{Kinematic Variables}

The kinematics of the interaction $ep\to\nu X$ are defined by the four--momenta
$k$ and $P$ of the incident electron and proton respectively, and the
four--momentum of the final state neutrino ($k'$) or the final state hadronic
system ($P'$). The four--momentum transfer between the electron and the proton
is given by $q=k-k'=P'-P$.

In addition to $s=(k+P)^2$, the square of the $ep$ center of mass energy, three
Lorentz invariants are defined in terms of these four--momenta:
\begin{itemize}
\item 
$\displaystyle Q^2=-q^2\;$, the negative square of the four--momentum transfer,
\item
$\displaystyle x={Q^2\over2q\cdot P}\;$, the Bjorken scaling variable,
\item
$\displaystyle y={q\cdot P\over k\cdot P}\;$, the fractional energy transfer
to the proton in its rest system.
\end{itemize}

\vskip-2.mm
\noindent
Neglecting mass terms, these variables are related by $Q^2=x\,y\,s$.

The Jacquet--Blondel method \cite{JB} was used to reconstruct the kinematic
variables from the measured missing transverse momentum,
\begin{equation}
  \ptmiss=\sqrt{\left(\sum_i p_X^i\right)^2+\left(\sum_i p_Y^i\right)^2}\;,
  \label{ptmissdef}
\end{equation}
and the quantity $\delta$, given by
\begin{equation}
  \delta=\sum_i\left(E^i-p_Z^i\right)\;,
  \label{empzdef}
\end{equation}
where the sums run over all EMC (HAC) cells with energy deposits $E^i$
above 60 MeV (110 MeV) and
the $\vec{p\,}^i$ are the momenta assigned to each calorimeter cell (calculated
assuming zero mass and using the geometrical cell center and the measured vertex
position). The Jacquet--Blondel estimators of $y$, $Q^2$ and $x$
are given by
\begin{equation}
  y_{JB}  ={\delta\over2E_e}\;,\qquad 
  Q^2_{JB}={\ptmiss^2\over1-y_{JB}}\;,\qquad
  x_{JB}  ={\ptmiss^2\over s\,y_{JB}(1-y_{JB})}\;,
  \label{JBdef}
\end{equation}
where $E_e$ is the electron beam energy.

\subsection{Selection of Charged--Current Events}

The number of CC triggers was about $10^5$, almost entirely due to $p$--gas
interactions, cosmic rays, and beam halo muons.  After a preselection which
required a tracking vertex and stricter timing cuts than at the trigger level,
about 6300 $e^+p$ and 1500 $e^-p$ candidates remained (cf.~table~\ref{cuteff}).
In the following, we describe the cuts which were applied to extract the CC
signal. Table \ref{cuteff} gives the numbers of events remaining after each
selection cut, as well as the corresponding efficiencies estimated using the
{\sc ariadne} MC samples.

The following conditions were imposed on all events passing the trigger:
\begin{itemize}
\item
$\ptmiss>11\,\gev$ was required. The trigger simulation indicated that this cut
was sufficiently far above the  $9\,\gev$ trigger threshold to 
ensure high trigger efficiency.
\item
The events had to have a tracking vertex with $|\zvtx|<45\,{\rm cm}$, where
$\zvtx=0$ at the nominal interaction point. This cut eliminated a large fraction
of non--$ep$ background.
\item
$\ptmiss^{\rm out}/\ptmiss>0.6$ was required, where $\ptmiss^{\rm out}$ is the
net transverse momentum for all calorimeter cells with a polar angle above
$9^\circ$ with respect to the nominal vertex position. This cut rejected
$p$--gas and $p$--beampipe collisions, for which {\ptmiss} is concentrated at
small polar angles.
\item
There had to be at least one reconstructed track originating from the vertex,
which had a polar angle between $15^\circ$ and $164^\circ$ and a transverse
momentum exceeding $0.2\,\gev$. This requirement removed cosmic ray events and
$p$--gas interactions with spurious vertices caused by low--energy secondary
interactions in the beam pipe.
\item
The difference $\Delta\phi$ between the azimuth of the net transverse momentum as
measured by the CTD tracks with polar angle between $15^\circ$ and $164^\circ$,
and the azimuth measured by the calorimeter, was required to fulfill
$|\Delta\phi|<1\,{\rm rad}$. This requirement removed overlays of cosmic rays on
$ep$--interactions.
\item
In addition, a pattern recognition algorithm based on the topology of the
calorimeter energy distribution was applied to reject cosmic rays and beam halo
muons.
\end{itemize}

Simulation of backgrounds due to NC DIS and photoproduction interactions showed
that such events passing the above cuts were concentrated at low {\ptmiss}.  The
same was found for the non--$ep$ background.  Hence the following additional
cuts were applied to events with $\ptmiss<30\,\gev$:
\begin{itemize}
\item
$y_{JB}<0.8$, which reduced NC DIS background.
\item
$\ptmiss/E_t>0.4$, where $E_t=\sum_i\sqrt{(p_X^i)^2+(p_Y^i)^2}$ is the total
transverse energy. This cut demanded an azimuthally collimated energy flow and
rejected photoproduction and also $p$--gas background.
\item
$P_t^{\rm tracks}/\ptmiss>0.1$, where $P_t^{\rm tracks}=\sqrt{\left(\sum_j
P_X^{{\rm track}\,j}\right)^2+\left(\sum_jP_Y^{{\rm track}\,j}\right)^2}$ and
$j$ runs over all vertex--fitted tracks with polar angle between $15^\circ$ and
$164^\circ$. This cut effectively tightened the $\ptmiss^{\rm out}/\ptmiss$
requirement (cf.~above) and also removed events with additional non--$ep$
related energy deposits in the calorimeter (mainly cosmic rays).
\end{itemize}

\begin{figure}[p]
  \centerline{\psfig{figure=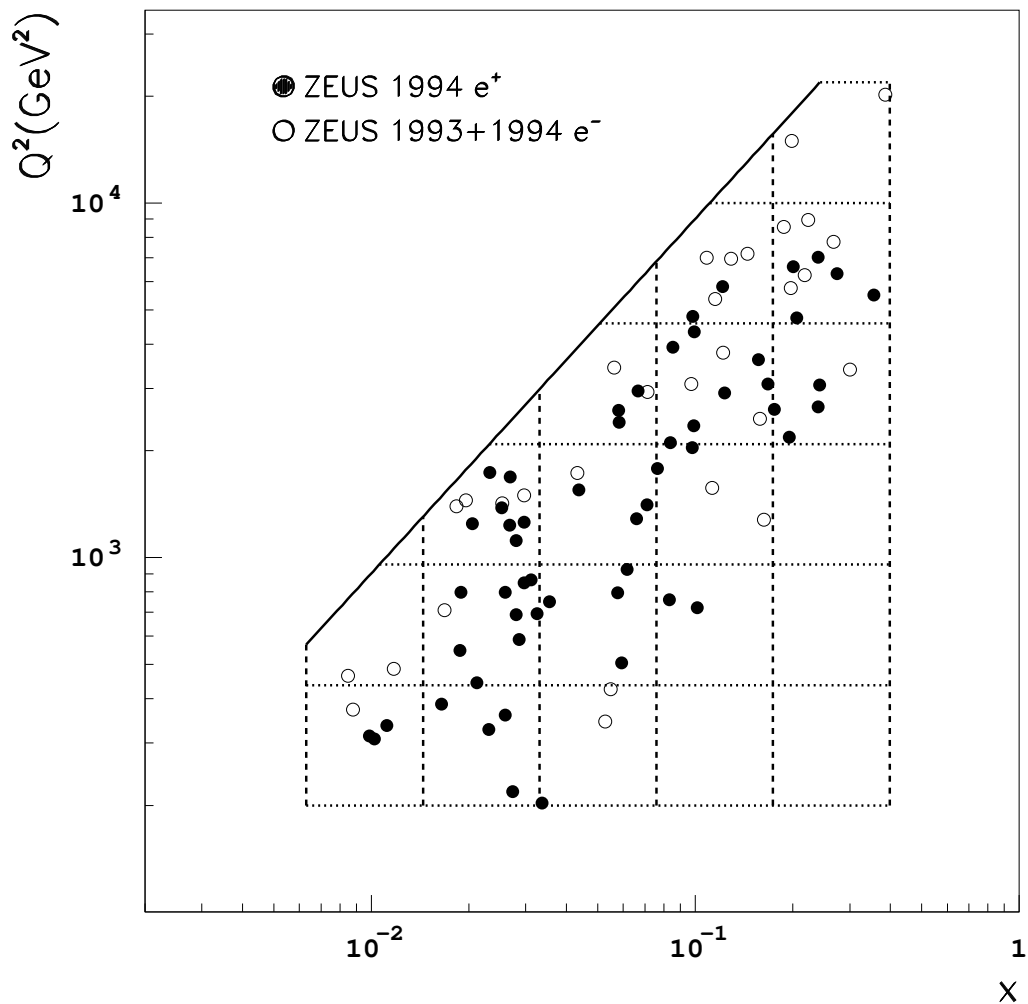,height=15cm}} 
  \caption{
Distribution of events in the $(x,Q^2)$ plane. The filled (open) circles
indicate the selected $e^+p$ ($e^-p$) CC events in a sample with integrated
luminosity of $2.93\,{\rm pb}^{-1}$ ($0.82\,{\rm pb}^{-1}$).  The horizontal
lines of the grid show the $Q^2$, the vertival lines the $x$ bins used for the
analysis. The solid diagonal line indicates the kinematic limit $y=1$.
  \label{xvsqsq} }
\end{figure}

The resulting sample was visually scanned and 3 non--$ep$ background events
were identified in the $e^+p$ sample and removed (all were overlays of muons and
$p$--gas reactions). No non--$ep$ background was found in the $e^-p$ sample.
The distribution in $x$ and $Q^2$ of the final samples of 56 $e^+p$ events and
30 $e^-p$ events is shown in figure~\ref{xvsqsq}.

\subsection{The Neutral--Current Control Sample}

The acceptance corrections and the methods of reconstructing the kinematic
variables were based on the MC simulation described in section~2.3.  In order to
verify the accuracy of the simulation, several systematic checks were performed
using a sample\footnote{Data collected in 1994 were used.} of high--$Q^2$ NC
events selected as follows:

\begin{itemize}
\item 
As in the CC case (cf.\ section~2.2), the NC trigger requirements were based on
the calorimeter energy deposits. The same timing cuts were applied as for the CC
trigger. Photoproduction background was efficiently suppressed by the cut
$\delta>25\,\gev$ (cf.\ eq.~\ref{empzdef}).
\item
In the offline NC selection, an electron candidate with at least $10\,\gev$
energy had to be identified by a neural--network algorithm using the pattern of
the calorimeter energy deposits \cite{sinistra}. The electron had to be isolated
(less than $2\,\gev$ of energy not associated with the electron in an
$(\eta,\phi)$--cone\footnote{The pseudorapidity $\eta$ is defined as
$-\ln[\tan(\theta/2)]$, where $\theta$ is the polar angle with respect to the
proton beam direction, taken from the reconstructed interaction point.} of
radius $R=0.5$ centered on the electron direction) and had to have a matching
track if it was in the region $20^\circ<\theta<160^\circ$. In addition, the
event had to have a reconstructed tracking vertex, and satisfy $35\,\gev<\delta<
65\,\gev$ and $Q^2>200\,\gev^2$, where $Q^2$ was reconstructed using the double
angle (DA) method \cite{DA}.
\item
For the selected NC events, the track(s) and the calorimeter deposits associated
with the electron were deleted from the data, and the reconstruction of the
event vertex was repeated.
\item
These modified NC events were then processed through the complete CC analysis
chain. In order to account for the different cross sections, a weight
\begin{displaymath}
  w={d\sigep/dx\,dQ^2\over d\sigepnc/dx\,dQ^2}\quad\hbox{for $e^+p$}\qquad
  \hbox{resp.}\qquad
  w={d\sigem/dx\,dQ^2\over d\sigemnc/dx\,dQ^2}\quad\hbox{for $e^-p$}
\end{displaymath}
was assigned to each event, where leading order differential cross sections,
evaluated at $x$ and $Q^2$ as reconstructed by the double angle method, were
used.

After the CC selection, 2269 $e^+p$ and 167 $e^-p$ events with $Q^2>200\,\gev^2$
remained, where $Q^2$ was reconstructed as described in section~4.1. The
respective sums of weights were $50.8$ and $6.2$.
\end{itemize}

The control sample was not corrected for the efficiency of the NC event
selection, for remaining backgrounds, or for migrations due to mismeasurement of
$x$ and $Q^2$. The distributions of $Q^2_{JB}$, $x_{JB}$, $y_{JB}$ and {\ptmiss}
agreed sufficiently well with the corresponding CC MC distributions to allow for
quantitative checks of the efficiency calculations and correction methods.

\section{Cross Section Evaluation}

The integrated and differential cross sections were calculated from the observed
numbers of events and the integrated luminosity after subtracting the estimated
photoproduction background and correcting for detection efficiencies and
migrations.

The $e^-p$ data from 1993 and 1994 were combined by appropriately weighting the
1993 event numbers in each individual bin under consideration. The weights,
which account for slight differences in the electron beam energies and in the
trigger and detector configurations, were in the range between $0.95$ and
$1.05$.

All statistical errors have been evaluated using the observed numbers of events
and asymmetric Poisson confidence intervals.

\subsection{Reconstruction of Kinematic Variables}

Due to the energy loss of the hadrons in inactive material in the central
detector, the calorimeter measurement underestimated both {\ptmiss} and
$\delta$.  The detector simulation was used to derive corrected values
$\ptmiss_{\,\rm cor}$ and $y_{\rm cor}$, as second--order polynomials of the
respective raw values {\ptmiss} and $y_{JB}$, where the coefficients for the $y$
correction depend on {\ptmiss}.  The correction of {\ptmiss} was about $20\%$ at
low {\ptmiss}, decreasing to $\approx10\%$ for $\ptmiss\gsim50\,\gev$; in $y$,
the corrections were highest for $y\approx0.1$ and low {\ptmiss}
($\approx20\%$), decreasing with both increasing $y$ and increasing {\ptmiss}.
The corrected values ${Q^2}_{\rm cor}$ and $x_{\rm cor}$ were calculated in
terms of $\ptmiss_{\,\rm cor}$ and $y_{\rm cor}$ using equation~\ref{JBdef}.

The correction was tested using the NC control sample (cf.\ section~3.3) by
evaluating the means and widths of the distributions of $\delta A=(A_{\rm
cor}-A_{DA})/A_{DA}$ for $A=Q^2,x$ or $y$ in all intervals of these variables
(cf.~section~4.2).  Typically, the mean values were below $10\%$ for $x$ and
$Q^2$ and below $5\%$ for $y$; the r.m.s.\ widths were roughly $20\%$ for $x$,
$25\%$ for $Q^2$, and $10\%$ for $y$. In the highest $x$ and $Q^2$ bins, these
estimates are based on only a few NC events and hence are of limited statistical
significance.

\subsection{Choice of Intervals}

The kinematic ranges of the variables $x$, $y$ and $Q^2$ covered by this
analysis are $x\greater0.006$, $0\less y\less1$ and $Q^2\greater200\,\gev^2$.
For the measurement of differential cross sections in $x$ and $Q^2$, equal bins
in $\log(x)$ and $\log(Q^2)$ respectively were used (see
tables~\ref{resQtotpos},~\ref{resQtotele} and~\ref{resx}).  The $y$ range was
subdivided in 5 equal intervals. These choices of bins provided roughly equal
numbers of events in all intervals and ensured that the intervals were much
wider than the experimental resolutions and the systematic shifts described in
section~4.1.

\subsection{Background Subtraction}

The contamination from NC events, estimated using a {\sc lepto} Monte Carlo
sample \cite{lepto}, was negligible.  The background from photoproduction
processes was determined by applying the selection algorithm to photoproduction
events, simulated with {\sc herwig}, which had transverse energy $E_t>20\,\gev$
($E_t$ being determined from the momenta of all final state particles).  While
the simulated photoproduction events which passed the CC selection all had
$E_t>30\,\gev$, they were concentrated at low {\ptmiss}.  The estimated
photoproduction contamination in the lowest bins of $x$ and $Q^2$ was
$\sim15\%$.

No event passing the CC selection was found in the MC samples simulating heavy
quark photoproduction, yielding an upper limit on the cross section for this
background of less than $0.2\%$ of the CC cross section. The heavy quark
background is therefore neglected.

No subtraction has been applied for non--$ep$ background (cosmic and beam halo
muons, $p$--gas reactions, and overlays of such events with $ep$ reactions).
This background is negligible after event selection and visual scanning (cf.\
section~3.2).

\subsection{Acceptance Correction and Unfolding}

Bin--to--bin migrations and the efficiencies of trigger and selection cuts were
taken into account using a bin--by--bin correction method, where the correction
factors were calculated from the Monte Carlo simulation described in section~2.
Typical acceptance corrections range from $40\%$ to $80\%$, with an average of
about $60\%$ (cf.\ tables~\ref{cuteff} and \ref{resQtotpos} to~\ref{resy}).  The
statistical errors of the correction factors were about 10~times smaller than
the statistical errors of the data and have been neglected. In addition to the
acceptance corrections, $\zeta$, defined as the ratio of the number of events
reconstructed in a given bin to the number of events generated in the same bin,
tables~\ref{resQtotpos} to~\ref{resy} also show the purities, $\xi$, defined as
the fraction of the events reconstructed in a given bin which were also
generated in the same bin.

The efficiency calculation was checked using the NC control sample.  Both the
overall acceptances and those of individual cuts (cf.~table~\ref{cuteff})
typically agreed with the MC values within $5\%$.

The bin--by--bin unfolding procedure described above was cross checked by
applying a single step matrix unfolding based on Bayes' theorem \cite{Julio}.
The off--diagonal elements of the unfolding matrix were usually less than $25\%$
of the corresponding diagonal elements, indicating that migrations between bins
were small. In cases where the data are poorly modeled by the MC or where
migrations play a dominant r\^ole, the two methods are expected to yield
significantly different results.  The resulting differences were however found
to be well below the statistical errors. They are included in the systematic
uncertainties (cf.\ tables~\ref{resQtotpos} and~\ref{resQtotele}).

\subsection{Systematic Uncertainties}

In the following, we summarize the studies which were performed in order to
estimate the systematic uncertainties. For the case of $d\sigma/dQ^2$, the
dominant systematic uncertainties are detailed in tables~\ref{resQtotpos} and
\ref{resQtotele}. The remaining systematic errors were all below $5\%$
and typically below $1\%$.  Where appropriate, independent checks of systematic
uncertainties have been calculated using the NC control sample. The systematic
deviations observed in this sample agreed well with the values quoted below.

\begin{itemize}
\item 
In the detector simulation, the calorimeter energy scale was varied by $\pm3\%$
($\pm5\%$ in the FCAL), corresponding to the level at which it is currently
understood. Using the BAC it was checked that this variation also covers
possible effects due to energy leakage out of the CAL. Using the BAC energy
deposits to correct jet energies in CC events increased {\ptmiss} by about $2\%$
on average.

Such an energy scale uncertainty gives rise to uncertainties in the cross
section which are typically between 5\% and 10\%, but can be up to $30\%$ in the
highest $Q^2$ and $y$ bins.
\item 
The systematic uncertainty of the photoproduction background subtraction was
estimated as the quadratic sum of its statistical error and an additional
$\pm50\%$ uncertainty which accounted for differences between $E_t$ and
$\ptmiss/E_t$ spectra of photoproduction data and MC.

The uncertainty of the background estimate mainly affected the lowest $Q^2$ and
$x$ bins. It was the dominant systematic error in the lowest $x$ bin for $e^+p$
reactions.
\item
The reconstruction of the kinematic variables was modified in the following
way: in each bin\footnote{For $\ptmiss$, the bin boundaries were $0\mns20\mns
30\mns40\mns50\mns\infty\,\gev$.} of $A=\ptmiss$ or $y$ the reconstructed values
of $\ptmiss$ and $y$ were multiplied with $1/(1+\langle\delta A\rangle)$
(cf.~section~4.1), where $\delta A$ was averaged over the weighted NC control
sample events in this bin.  This corresponded to correcting on average to the DA
values as derived from the NC control sample.  $Q^2$, $x$ and $y$ were
calculated from the resulting values using equations~\ref{JBdef}. This changed
the measured cross sections by $10\%$ or less, except in the lowest and highest
$Q^2$ bins, where variations of up to $15\%$ were seen.
\item
In order to check the simulation of the tracking, the combined acceptance of all
cuts using tracking information was evaluated for the NC control sample, in each
interval of $Q^2$, $x$ and $y$, both for data and for simulated NC events.  The
data/MC ratio of these acceptances was found to be consistent with unity.  In
almost all bins the deviation was less than $3\%$, which was used as the overall
systematic uncertainty.
\item 
A one step Bayes unfolding was used instead of the bin--by--bin unfolding
(cf.~discussion in section~4.4). This variation caused uncertainties of
$10-15\%$ in the lowest and the highest bins of $Q^2$, $x$ and $y$ ($-25\%$ in
the highest $y$ bin for $e^+p$).
\item 
In the trigger simulation, the first level trigger thresholds were increased to
a level where the measured trigger efficiencies were close to $100\%$.  Except
in the lowest $Q^2$ bins, where effects up to $4\%$ were observed, this had a
negligible effect indicating that the efficiency calculations were insensitive
to the detailed shape of the trigger turn--on curves.
\item
In order to test the sensitivity to the details of the parton shower simulation,
the {\sc meps} model was used instead of the {\sc ariadne} MC for calculating
acceptance corrections and for unfolding. This affected the cross section
results only negligibly.

In addition, the selection cuts on $\ptmiss$, $\Delta\phi$, the track
requirements, and the cuts at low $\ptmiss$ have been varied using the {\sc
meps} MC sample instead of the data, and the {\sc ariadne} sets for acceptance
corrections and unfolding. This tested whether the simulated shapes of the
relevant distributions differ in the vicinities of these cuts.

Each of the cut variations changed the results by less than typically $1\%$,
except the $\ptmiss$ variation in the lowest $x$ and $Q^2$ bins (up to $5\%$)
and the track requirement (up to $3\%$ in some bins).
\item
Different parton distribution functions (MRSD0 \cite{MRSD0}, MRSD$_{-}'$
\cite{MRSDmnsp}, GRV \cite{GRV} and CTEQ~2pM \cite{CTEQ}) were used in the MC
samples used for the acceptance corrections.\footnote{The most significant
variation of the Standard Model cross section prediction due to parton
distributions is seen in $d\sigma/dx$ at $0.01\lsim x\lsim0.1$
(cf.~section~5.1).} The resulting variations were small and have been neglected.
\end{itemize}

The overall systematic uncertainty for each result (including ratios of cross
sections and the $W$ mass) was calculated by evaluating the variation of this
result due to each of the modifications of the analysis procedure described
above (including the luminosity uncertainties of $\pm2.0\%$ for the $e^+p$
sample and $\pm3.0\%$ for the $e^-p$ sample) and then separately summing in
quadrature the positive and negative deviations.

\section{Results}

The results will be presented in three sections: cross sections and the
determination of $M_W$, jet rates and profiles, and finally the search for
events with a large rapidity gap. The abbreviation `SM' (for Standard Model) is
used to indicate the theoretical predictions.  Unless otherwise noted, these
were estimated using {\sc lepto} and the MRSA parton distributions. The
propagator four--momentum was used to calculate kinematic quantitites.

\subsection{Cross Sections and Cross Section Ratios}

\vskip3.mm
\leftline{\uline{Differential cross sections as functions of $Q^2$, $x$ and $y$}}
\vskip2.mm

\begin{figure}[p]
  \centerline{\psfig{figure=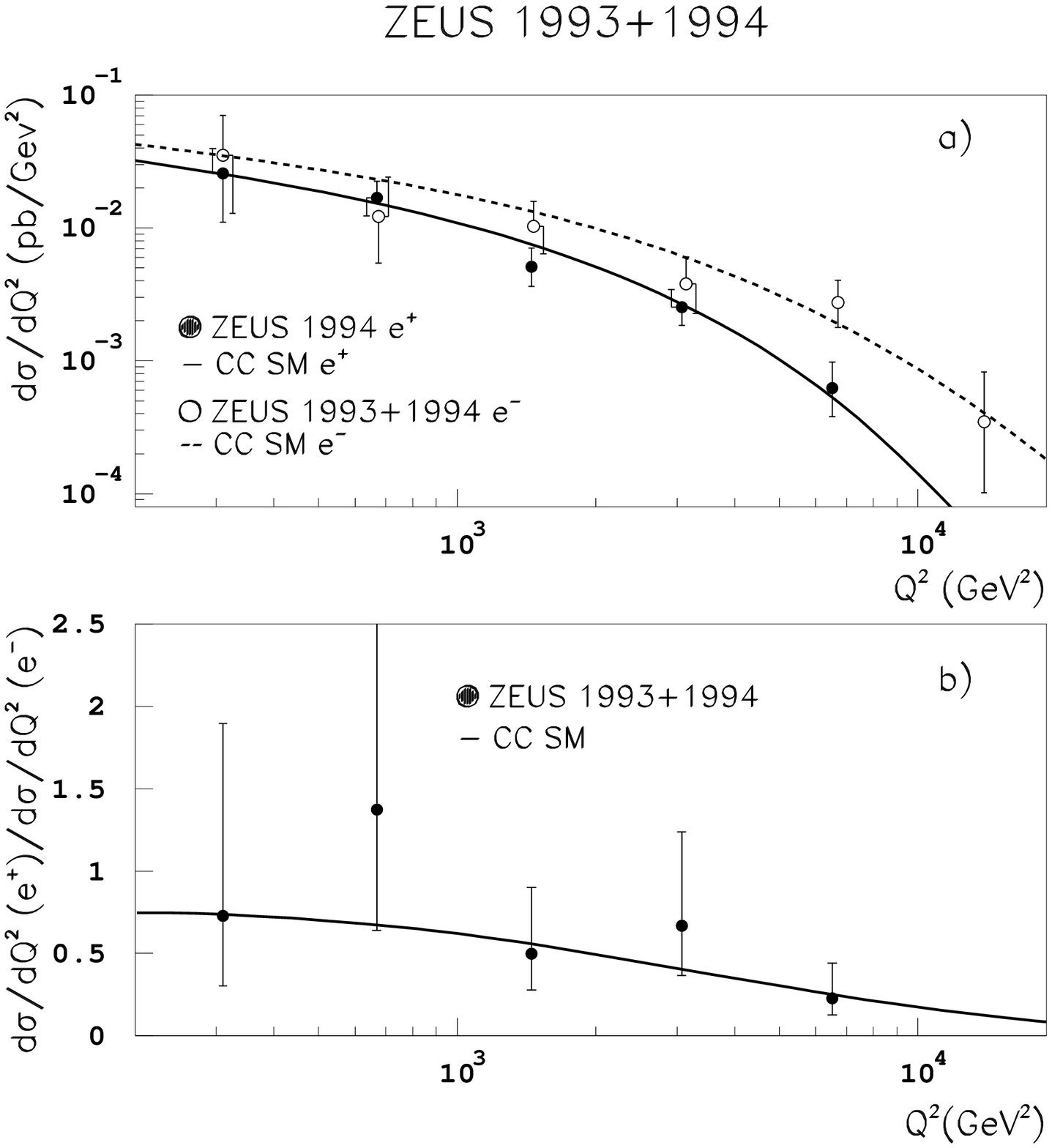,height=18cm}}
  \caption{
Differential CC cross section versus $Q^2$. In a), the filled (open) dots and
the solid (dashed) curve represent the measured values and the Standard Model
(SM) prediction of $d\sigma/d\,Q^2$ for $e^+p$ ($e^-p$) collisions. Plot b)
shows $\sigrat{Q^2}$. The error bars indicate the statistical and systematic
uncertainties combined in quadrature.  The horizontal position of the points is
given by the generator level $Q^2$ average of the MC events in each bin.
  \label{dsdq} 
}
\end{figure}

The values of the cross section, $\sigma_i$, in bins of $Q^2$, for $e^+p$ and
$e^-p$ data, are listed in tables \ref{resQtotpos} and \ref{resQtotele}
respectively. Also shown are the dominant systematic errors as described in
section~4.5.  The results are plotted in figure~\ref{dsdq}a, where the SM
predictions are also shown.

\begin{figure}[p]
  \centerline{\psfig{figure=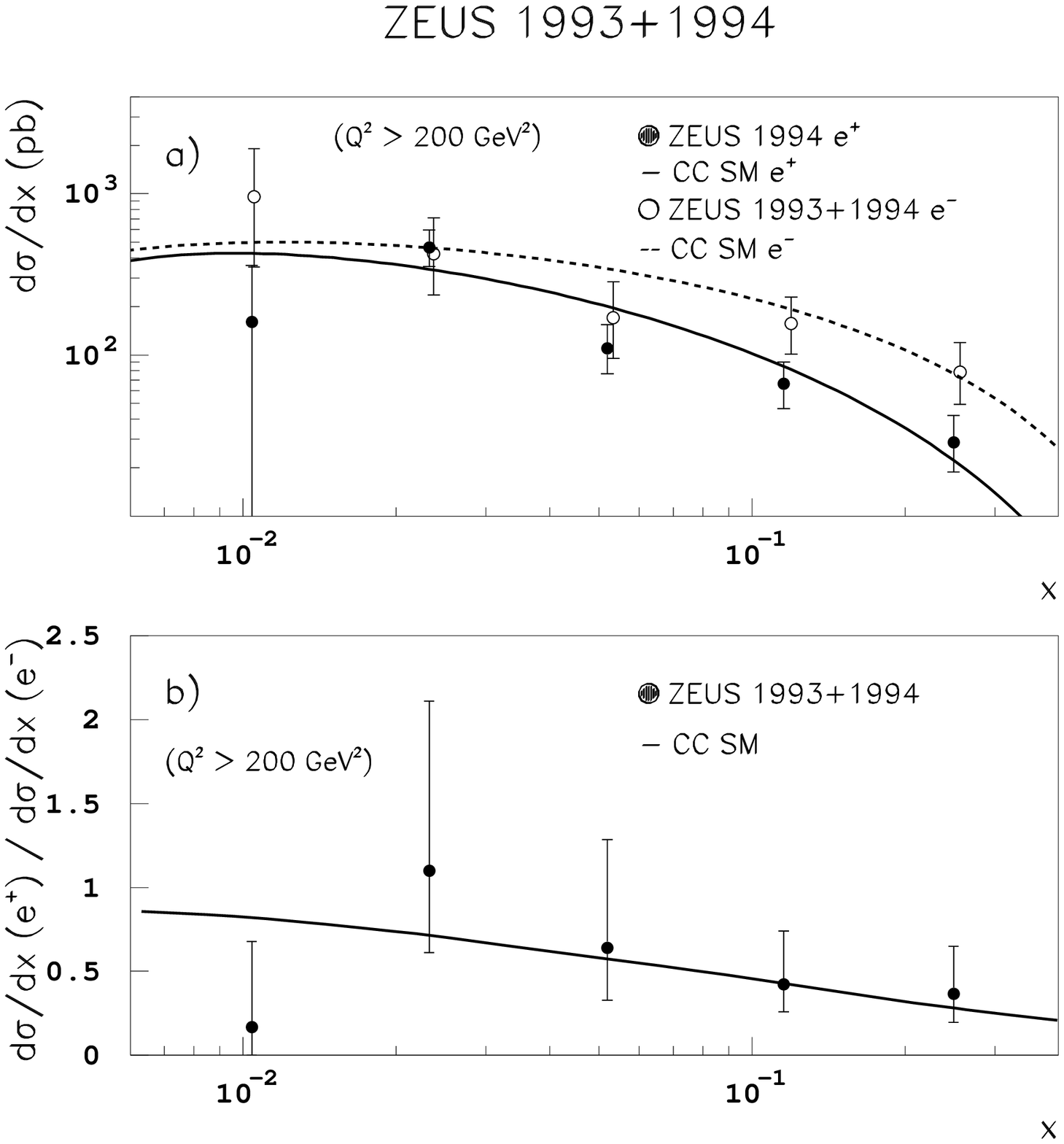,height=18cm}}
  \caption{
Differential CC cross section versus $x$ for $Q^2>200\,\gev^2$. In a), the
filled (open) dots and the solid (dashed) curve represent the measured values
and the Standard Model (SM) prediction of $d\sigma/d\,x$ for $e^+p$ ($e^-p$)
collisions. Plot b) shows $\sigrat{x}$. The error bars indicate the statistical
and systematic uncertainties combined in quadrature.  The horizontal position of
the points is given by the generator level $x$ average of the MC events in each
bin.
  \label{dsdx}
}
\end{figure}

\begin{figure}[p]
  \centerline{\psfig{figure=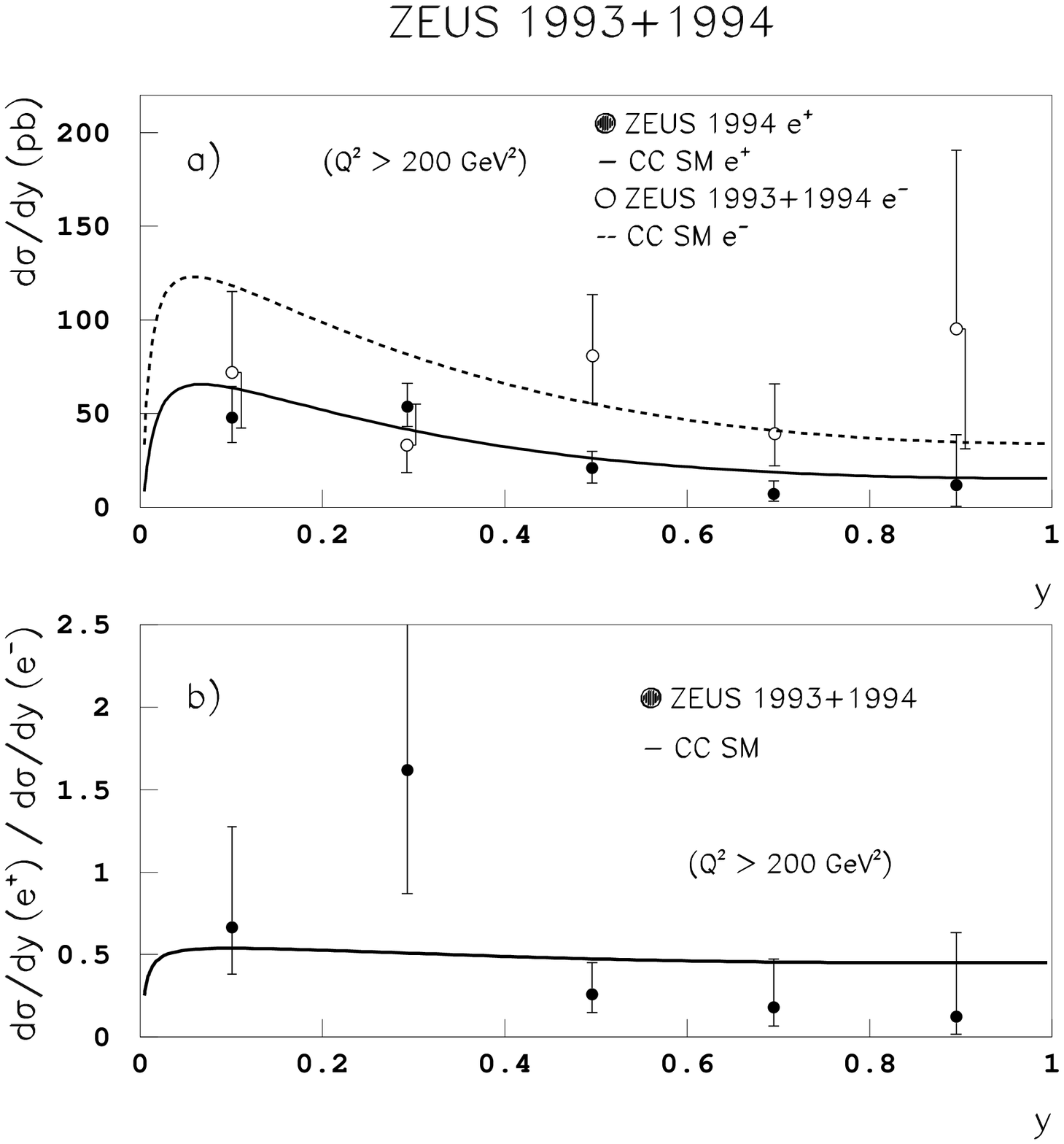,height=18cm}}
  \caption{
Differential CC cross section versus $y$ for $Q^2>200\,\gev^2$. In a), the
filled (open) dots and the solid (dashed) curve represent the measured values
and the Standard Model (SM) prediction of $d\sigma/d\,y$ for $e^+p$ ($e^-p$)
collisions. Plot b) shows $\sigrat{y}$. The error bars indicate the statistical
and systematic uncertainties combined in quadrature.  The horizontal position of
the points is given by the generator level $y$ average of the MC events in each
bin.
  \label{dsdy}
}
\end{figure}

The results for $Q^2\greater200\,\gev^2$ in bins of $x$ and $y$ are given in
tables~\ref{resx} and~\ref{resy}. The differential CC cross sections are shown
in figures~\ref{dsdx}a and~\ref{dsdy}a together with the corresponding SM
predictions.

Figures~\ref{dsdq}b, \ref{dsdx}b and~\ref{dsdy}b show the ratios $\sigep/\sigem$
as functions of the respective variables. The curves again represent the SM
predictions.

Several observations can be made:
\begin{itemize}
\item
No significant deviations of the measured cross sections from the SM predictions
are observed. This is quantified by the $\chi^2$ values which are listed in
table~\ref{tabchi}.
\item
The differential cross sections $d\sigma/d\,Q^2$ fall steeply with $Q^2$ (cf.\
figure~\ref{dsdq}), reflecting the influence of the $W$ propagator, the decrease
of the parton densities with increasing $x=Q^2/sy$, and the $(1-y)^2$ terms in
the cross section. The differences between the $e^+p$ and $e^-p$ cross sections
are due to the different $x$ and $y$ behavior (cf.\ discussion below).
\item
As can be seen from figure~\ref{dsdx}, at low $x$, $d\sigma/d\,x$ is about the
same for $e^+p$ and $e^-p$ scattering. This reflects the fact that the cross
sections (equations \ref{Opos},\ref{Oele}) become equal if valence quark
distributions can be neglected and if $\,\dbar(x,Q^2)=\ubar(x,Q^2)$.

The decrease of $d\sigma/d\,x$ with increasing $x$ is more rapid for $e^+p$
scattering than for $e^-p$ scattering. This behavior is also expected from
equations \ref{Opos} and \ref{Oele}: in $e^+p$ reactions, scattering on valence
quarks is reduced by the $(1-y)^2$ factor; in addition, the $e^+$ couples to
$d(x,Q^2)$, which is suppressed at high $x$ relative to $u(x,Q^2)$ which is
relevant for $e^-p$ scattering (cf.\ e.g.\ \cite{BEBCdoveru}).

The $x$ dependence of the $e^+p$ to the $e^-p$ cross section ratio is
illustrated by the ratios for $x<0.1$ and for $x>0.1$ (with $Q^2>200$ GeV$^2$):
\begin{displaymath}
  {\sigep\over\sigem}=
  \cases{
   {\displaystyle
   0.91\,{}^{+0.41}_{-0.27}\,{\rm [stat]}\pm0.04\,{\rm [syst]}}&
   \quad{\rm for}\quad\stru{2.ex}{2.8ex}$0<x<0.1$\cr
   {\displaystyle
   0.26\,{}^{+0.14}_{-0.09}\,{\rm [stat]}\pm0.01\,{\rm [syst]}}&
   \quad{\rm for}\quad\stru{2.8ex}{2.ex}$0.1<x<1\;.$}
\end{displaymath}
The corresponding SM predictions are $0.64$ and $0.32$.  The measured ratios
confirm the suppression of the $e^+p$ CC cross section at high $x$.

The SM predictions for $d\sigma/d\,x$ have also been evaluated using different
parton distribution functions (MRSD$_{-}'$, GRV, CTEQ~2pM).  The largest
differences have been observed between the sets MRSD$_{-}'$ and CTEQ~2pM for $x$
between 0.01 and 0.1, where the cross section predicted by MRSD$_{-}'$ is up to
$10\%$ lower than the CTEQ~2pM values, both for $e^+p$ and $e^-p$ scattering.
The MRSA, CTEQ~2pM and GRV predictions agree to better than $5\%$.

\item
The measured differential cross sections $d\sigma/d\,y$ agree with the
theoretical predictions. The SM curves in figure~\ref{dsdy} reveal that the
expected shapes are very similar for $e^+p$ and $e^-p$ scattering, indicating
that the relative contributions of the $(1-y)^2$ terms are about equal in both
cases.

As $y\to1$, the cross section ratio $e^+p/e^-p$ is given by $1/\left(
{\displaystyle u_v\over\displaystyle u_s+c}+1\right)$ (assuming $\ubar=u_s$ and
$\cbar=c$, the indices $v$ and $s$ indicating valence and sea quark
distributions, respectively). The fact that the prediction, corresponding to
$u_v/(u_s+c) \approx1$, is above the data for $y>0.4$, is only moderately
significant in view of the large errors.

At $y\to0$, the cross section ratio $e^+p/e^-p$ approaches $(d_v+S)/(u_v+S)$,
where $S=d_s+u_s+s+c$ (assuming again that the quark and antiquark sea
distributions are identical for each flavor).
\end{itemize}

\begin{figure}[p]
  \centerline{\psfig{figure=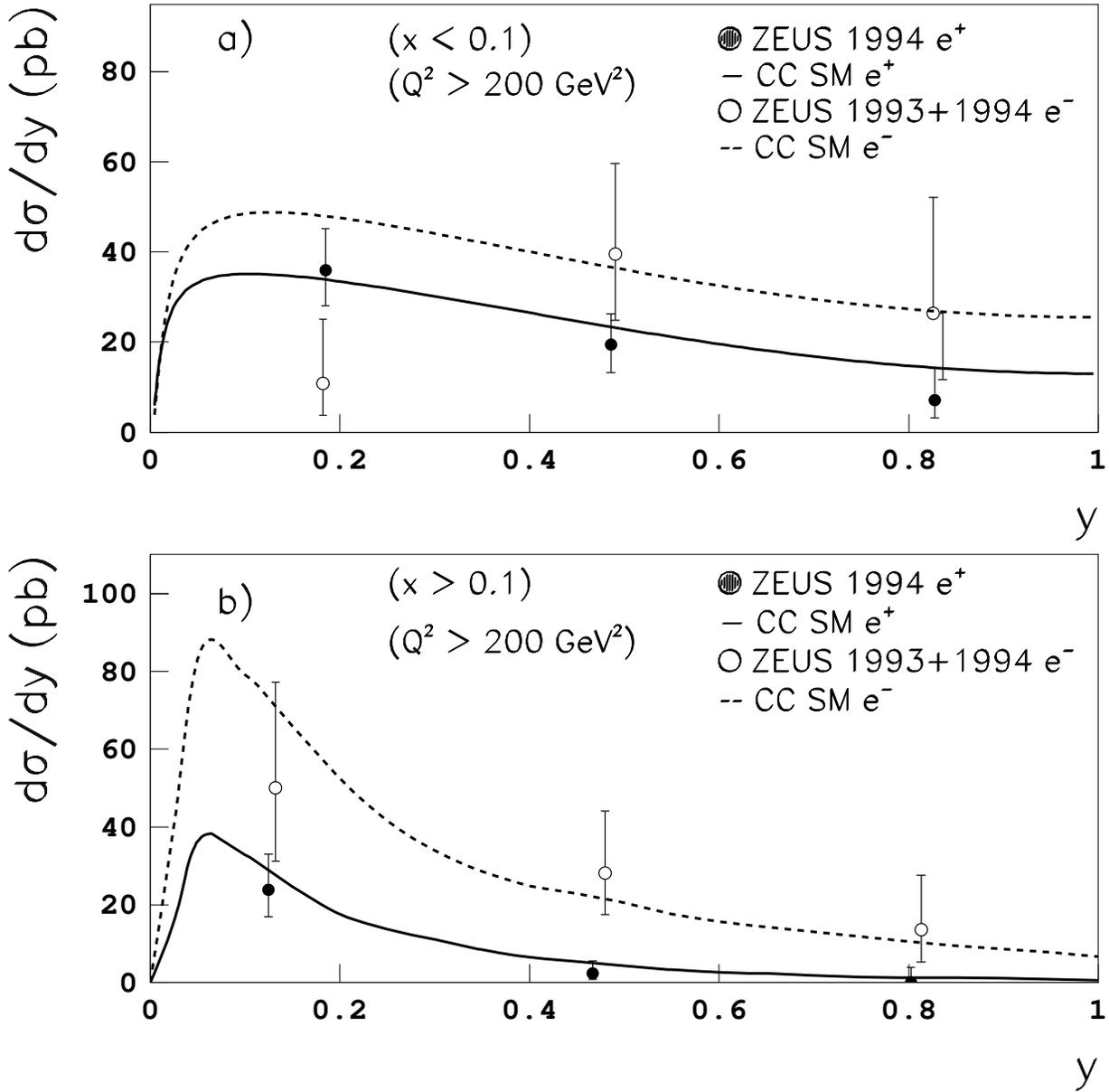,height=18cm}}
  \caption{
Differential CC cross section versus $y$ for $Q^2>200\,\gev^2$, a) for $x<0.1$
and b) for $x>0.1$. The filled (open) dots and the solid (dashed) curves
represent the measured values and the Standard Model (SM) prediction of
$d\sigma/d\,y$ for $e^+p$ ($e^-p$) collisions.  The error bars indicate the
statistical and systematic uncertainties combined in quadrature.  The horizontal
position of the points is given by the generator level $y$ average of the MC
events in each bin.
  \label{dsdy_x}
}
\end{figure}

In order to further study the influence of the $V\mns A$ helicity structure and
the parton distributions, figure~\ref{dsdy_x} shows $d\sigma/d\,y$ separately
for $x<0.1$ (where sea quarks dominate the cross section) and for $x>0.1$ (where
the valence contribution is larger). Here three equal size $y$ bins are used.

For low $x$, the predicted $y$ shapes are similar for $e^+p$ and $e^-p$
scattering, the differences being mainly due to residual valence contributions.
The most significant deviation of the measured cross sections from the
prediction is observed in the lowest $y$ bin ($y<0.33$), where the measured
$\sigem$ is about two standard deviations below the SM prediction.

For $x>0.1$, both the $e^+p$ and the $e^-p$ cross sections decrease with
increasing $y$.  For $e^-p$, this can be attributed to residual sea quark
contributions and to the $W$ propagator which enters via the dependence of the
average $Q^2$ on $y$.  For $e^+p$ scattering, the decrease of $d\sigma/d\,y$
with increasing $y$ is even faster due to the $(1-y)^2$ term in the cross
section. All measured values are compatible with the predictions, indicating
that the slight deviations observed at $y\sim 0.3$ in the $e^+p/e^-p$ ratios are
due to the low--$x$ data.

\newpage
\leftline{\uline{Integrated cross sections for $Q^2>200\,\gev^2$}}
\vskip2.mm

The integrated cross sections for $Q^2>200\,\gev^2$ are obtained by summing the
cross sections in the $Q^2$ bins of tables \ref{resQtotpos} and
\ref{resQtotele}, resulting in\footnote{Summing the single cross sections
implies using an average of the acceptances evaluated for the single bins, which
is slightly different from the global acceptance quoted in table~\ref{cuteff}.}
\begin{eqnarray*}
 \sigep(Q^2>200\,\gev^2)&=&
 30.3\,{}^{\pls\chax5.5}_{\mns\chax4.2}\,{\rm [stat]}\;
           {}^{\pls1.6}_{\mns2.6}\,{\rm [syst]}\,{\rm pb}\\
 \sigem(Q^2>200\,\gev^2)&=&
 54.7\,{}^{\pls15.9}_{\mns\chax9.8}\,{\rm [stat]}\;
           {}^{\pls2.8}_{\mns3.4}\,{\rm [syst]}\,{\rm pb}\;.
\end{eqnarray*}

\noindent
These measurements are compatible with the respective SM predictions of
$32.3\,{\rm pb}$ and $65.8\,{\rm pb}$. The $e^+p/e^-p$ ratio is

\begin{equation}
  {\sigep\over\sigem}(Q^2>200\,\gev^2)=
   0.55\,{}^{+0.16}_{-0.12}\,{\rm[stat]}\;{}^{+0.02}_{-0.03}\,{\rm[syst]}\;,
\end{equation}

\noindent
in agreement with the SM prediction of $0.49$.

\vskip3.mm
\leftline{\uline{Determination of the $W$ mass}}
\vskip2.mm

The $Q^2$ dependence of the CC cross section is largely determined by the
propagator term ${\cal P}(M_W,Q^2)\eql[M_W^2/(M_W^2+Q^2)]^2$. In order to
determine the value of $M_W$, the differential cross section was factorized
according to $d\sigma/dQ^2={\cal P}(M_W,Q^2)\cdot\Phi(Q^2)$. The function
$\Phi(Q^2)$, containing the $Q^2$ dependence of the parton densities and
couplings, was taken from the MC simulation, using the MRSA parton
distributions. A binned log--likelihood fit with $M_W$ as the free parameter,
applied simultaneously to the $e^+p$ and $e^-p$ data samples, yielded
\begin{equation}
  M_W=79\,{}^{+8}_{-7}\,{\rm [stat]}\;{}^{+4}_{-4}\,{\rm [syst]}\,\gev \;,
\end{equation}
in agreement with the average value $M_W=80.22\pm0.26\,\gev$ \cite{PDG95}
obtained from direct measurements of the $W$ mass at $p\overline p$ colliders
\cite{mwppbar} and also with the recent H1 measurement \cite{H1CC94Q}.  The
systematic uncertainty of $M_W$ has been evaluated using the method described
in section~4.5. The dominant contribution is the calorimeter energy scale.

\subsection{Jet Analysis}

An $(\eta,\phi)$--cone jet finding algorithm \cite{snowmass} has been applied to
the data, using a cone radius, $R=(\Delta\phi^2+\Delta\eta^2)^{\frac{1}{2}}$ of
$0.7$.  Pre--clusters are formed around calorimeter cells with transverse
energies larger than $0.3\,\gev$, and the final clusters are called jets if
their transverse energies exceed $6\,\gev$ and their pseudorapidities
$\eta^{\,\rm jet}$ are less than $2.5$ (i.e.\ polar angles greater than
$9^{\circ}$).  This jet analysis is done in the laboratory system. A detailed
analysis of jets in NC reactions, covering the region $Q^2\lsim10^3\,\gev^2$,
has previously been published in \cite{ZEUSjets}.

The results of the present jet analysis turn out to be insensitive to the choice
of the cone radius ($R=1$ and $R=0.5$ have also been tested). The constraint on
$\eta^{\,\rm jet}$ was applied in order to avoid effects from the fragmentation of
the proton remnant. In fact, the proton remnant manifests itself as a moderate
enhancement of the transverse energy flow around the forward edge of acceptance,
and for $\eta^{\,\rm jet}=2.5$ this enhancement is still clearly separated from
the $E_t$ flow associated with the jet.

\begin{figure}[h]
  \centerline{\psfig{figure=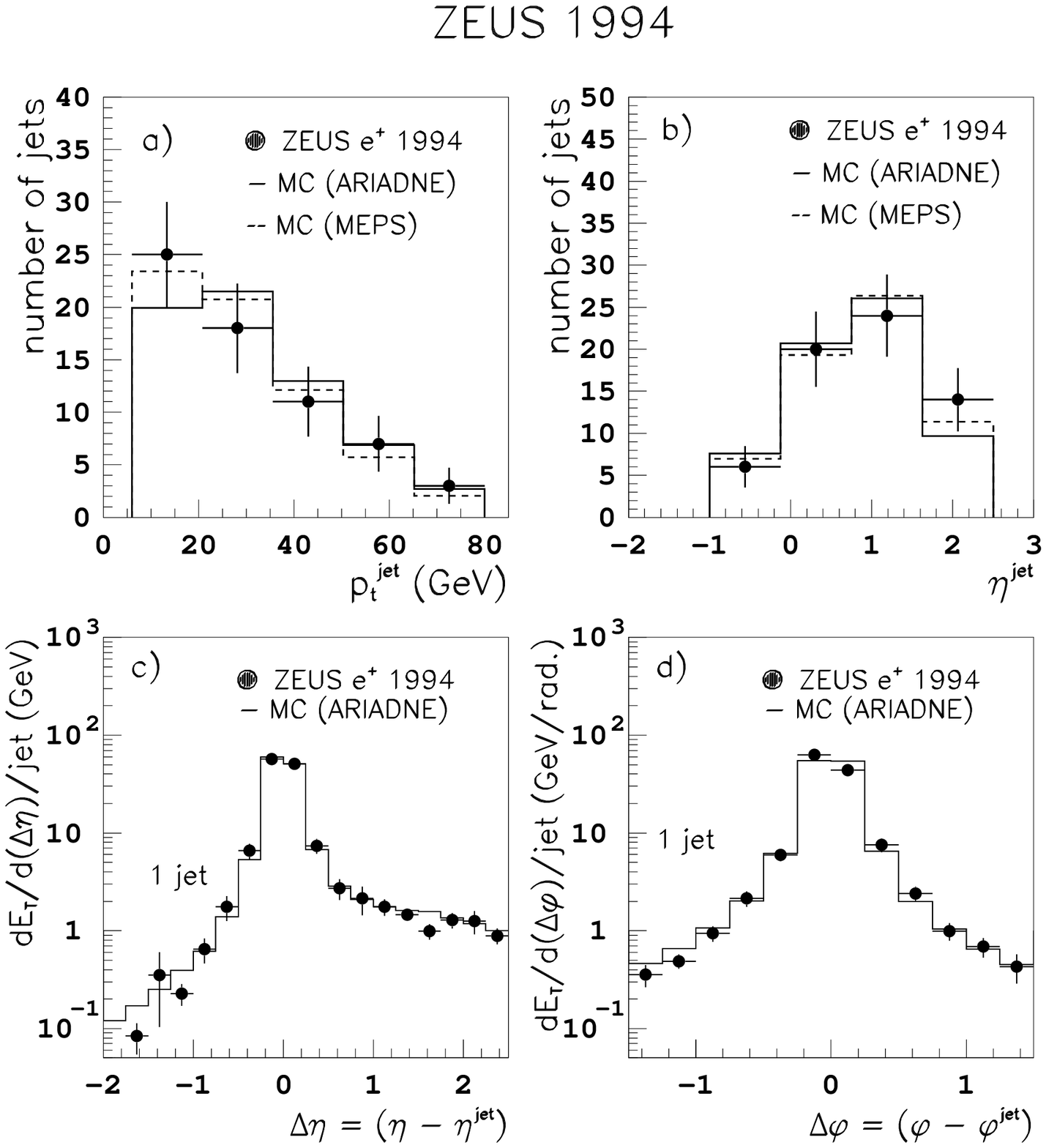,height=15cm}}
  \caption{
Plots a) and b) show the distributions of transverse momenta and
pseudorapidities of all jets found in the $e^+p$ sample.  Plots c) and d) give
the transverse energy profiles in $\eta$ and $\phi$ for jets in events with
exactly one jet.  Positive $\Delta\eta$ values correspond to angles between the
jet axis and the forward direction.  The points denote the data and the solid
(dashed) histograms represent the {\sc ariadne} ({\sc meps}) Monte Carlo
predictions including detector effects. In plots a) and b) the MC predictions
have been normalized to the number of jets observed in the data.
  \label{jetspos}
}
\end{figure}

\begin{figure}[h]
  \centerline{\psfig{figure=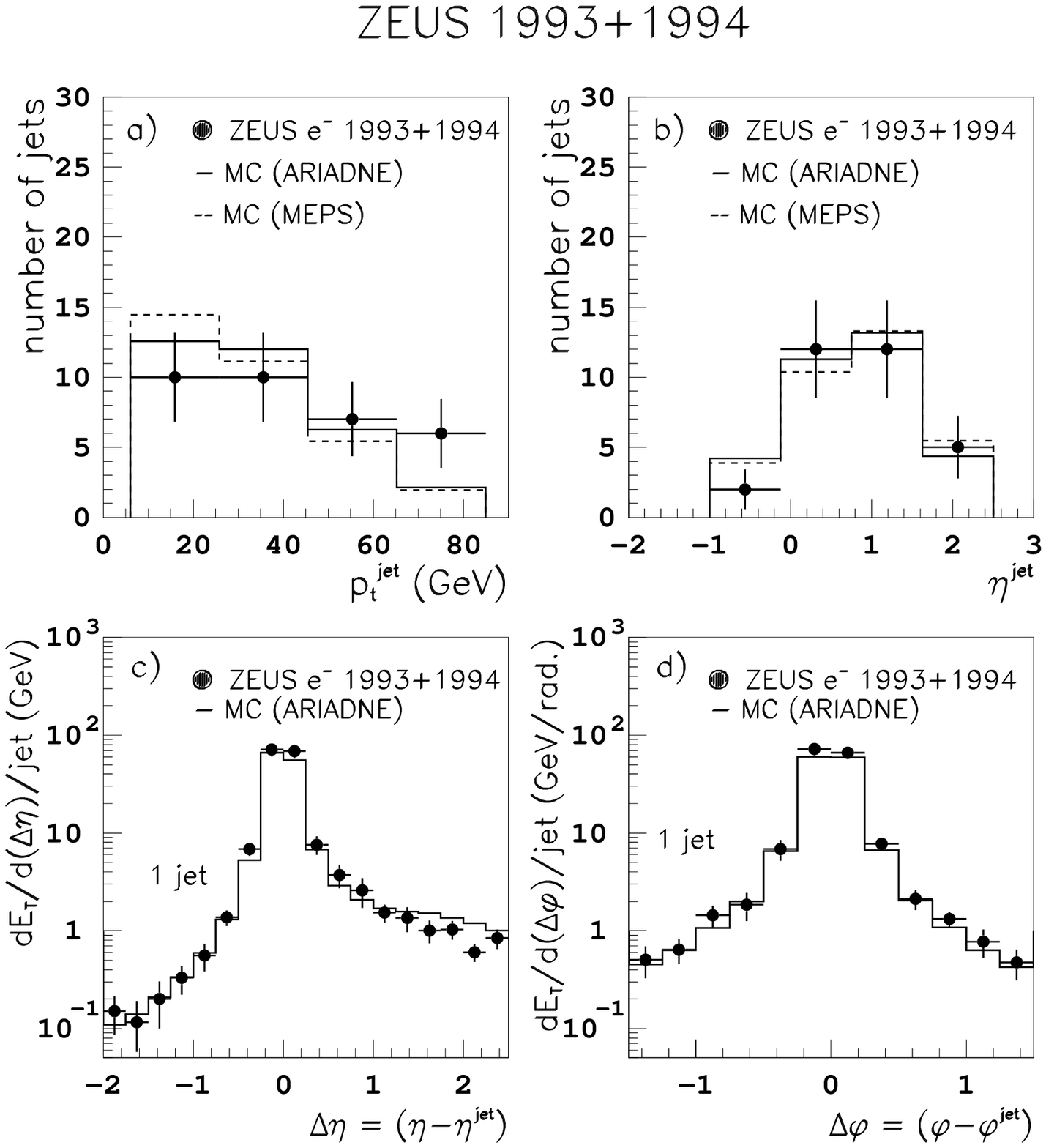,height=15cm}}
  \caption{
Plots a) and b) show the distributions of transverse momenta and
pseudorapidities of all jets found in the $e^-p$ sample.  Plots c) and d) give
the transverse energy profiles in $\eta$ and $\phi$ for jets in events with
exactly one jet.  Positive $\Delta\eta$ values correspond to angles between the
jet axis and the forward direction. The points denote the data and the solid
(dashed) histograms represent the {\sc ariadne} ({\sc meps}) Monte Carlo
predictions including detector effects. In plots a) and b) the MC predictions
have been normalized to the number of jets observed in the data.
  \label{jetsele}
}
\end{figure}

For all jets found, the distributions of the jet momentum transverse to the beam
axis, $p_t^{\,\rm jet}$, is shown in figures \ref{jetspos}a and \ref{jetsele}a
for $e^+p$ and $e^-p$ data, respectively. $p_t^{\,\rm jet}$ has been corrected
for energy loss in the inactive material of the detector\footnote{This was the
only correction applied in this jet analysis.} by using the average difference
of measured and true values of $p_t^{\,\rm jet}$ as determined in the MC
simulation as a function of the measured $p_t^{\,\rm jet}$.  Note that the
$p_t^{\,\rm jet}$ spectrum is much harder for the $e^-p$ than for the $e^+p$
data, as is expected from the harder $Q^2$ distributions. Figures \ref{jetspos}b
and \ref{jetsele}b show the distributions of $\eta^{\,\rm jet}$.  The transverse
energy flows measured in pseudorapidity ($\Delta\eta$) and azimuthal angle
($\Delta\phi$) relative to the jet axis, are plotted in figures \ref{jetspos}c,d
and \ref{jetsele}c,d. The hadronic energy flow produced between the jet and the
proton remnant is clearly observed in the excess of transverse energy at high
$\Delta\eta$, i.e.\ towards the proton remnant.

All these figures also show the predictions of the {\sc ariadne} MC. The
$p_t^{\,\rm jet}$ and $\eta^{\,\rm jet}$ distributions are also compared to the
{\sc meps} simulation. The MC distributions have been normalized to the number
of jets observed in the data.  For the $E_t$ flow distributions, the differences
between {\sc ariadne} and {\sc meps} (not shown) are slight. Good agreement is
observed between the data and both the {\sc ariadne} and the {\sc meps}
predictions, indicating in particular that the jet properties are well described
by the MC models.

The jet multiplicity distributions are summarized in table \ref{jetst}. All
selected events have at least one jet. Events with two or more jets are expected
from hard QCD subprocesses, in particular from boson--gluon fusion (BGF). The MC
predictions for the jet multiplicities, which include the ${\cal O}(\alpha_s)$
matrix elements, exhibit significant differences, amounting to nearly a factor
of two in the 2--jet rates. However, within their statistical errors, the data
are compatible with both models, favoring values intermediate between {\sc
ariadne} and {\sc meps}.

\subsection{Search for Large Rapidity Gap Events}

Events with a large rapidity gap (LRG) between the outgoing proton system and
the rest of the hadronic final state have been observed at HERA in
neutral--current DIS \cite{lrgdis}.  These events are generally understood to be
of a diffractive nature and to result from the exchange of a colorless object
(usually called the pomeron) with the quantum numbers of the vacuum.  Evidence
for a partonic structure of the pomeron in DIS has also been observed by the
HERA experiments \cite{lrgquarks}.

In CC processes, the coupling of the exchanged $W$ is sensitive to the flavor of
the pomeron constituents, which could provide additional information on the
pomeron structure. The search also is sensitive to high $Q^2$ diffractive
production of exclusive hadronic states such as vector or axial--vector mesons,
complementing corresponding NC studies \cite{DISVecMes}.

Similar to the prescription used in a previous NC analysis \cite{ZEUSDiffSF},
LRG events are identified using two variables, $\eta_{\rm max}$ and $\theta_H$.
$\eta_{\rm max}$ is defined as the maximum pseudorapidity of any calorimeter
cluster\footnote{Calorimeter clusters are groups of adjacent cells which have
energy deposits above the noise thresholds.} or cell with energy greater than
$400\,\mev$. The global quantity $\theta_H$ is given by the energy weighted mean
polar angle of the energy deposits in the calorimeter, $\cos\theta_H=\sum_ip_Z^i
/\sum_iE^i$, where the sums run over all calorimeter cells. In the na\"\i ve
quark parton model, $\theta_H$ is the scattering angle of the struck quark.

\begin{figure}[p]  
  \centerline{\psfig{figure=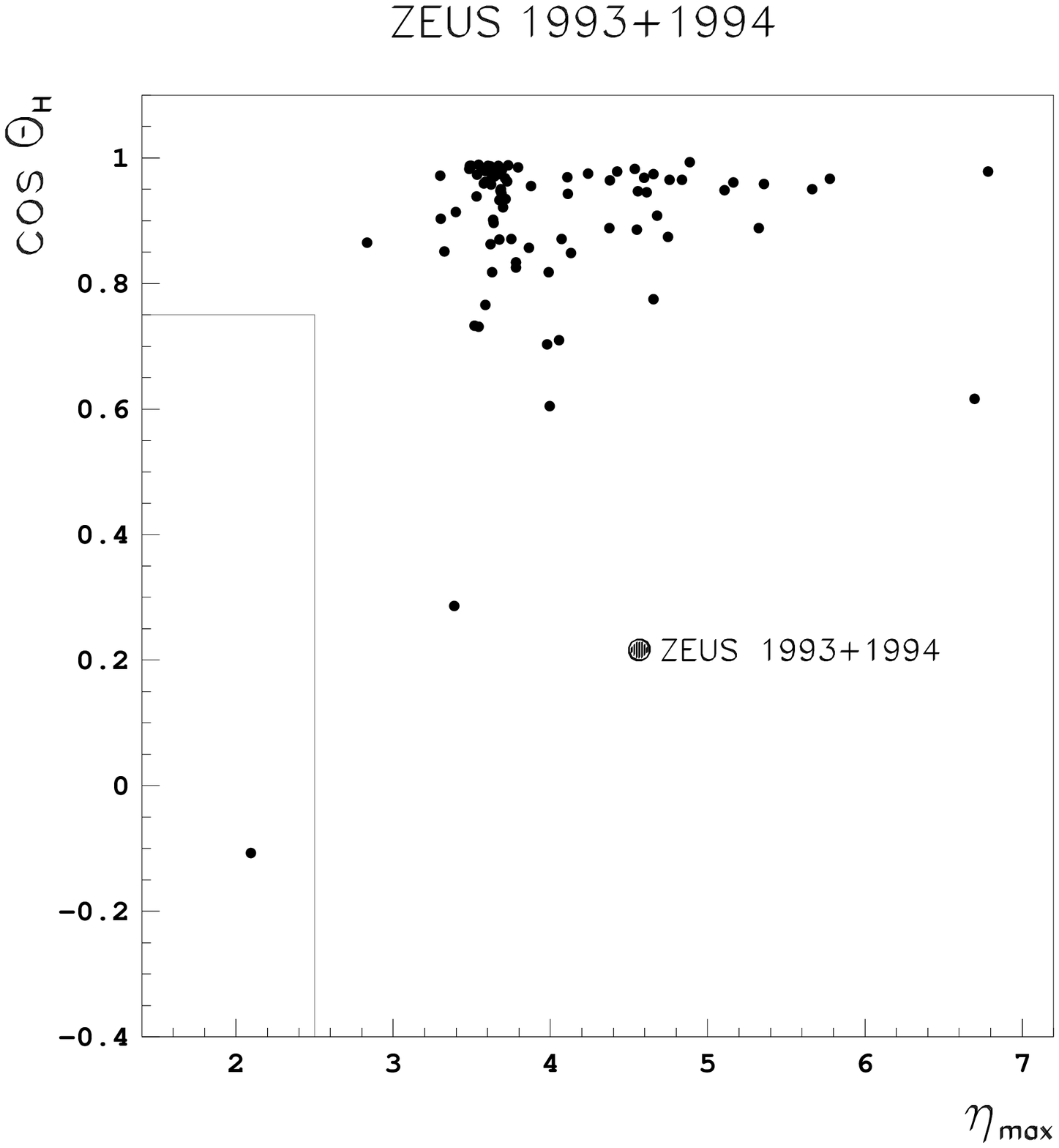,height=15cm}}
  \caption{
Scatter plot of $\cos\theta_H$ vs.\ $\eta_{\rm max}$ containing all events in
both the $e^+p$ and the $e^-p$ samples (see section~5.3 for the definitions of
$\theta_H$ and $\eta_{\rm max}$). The region $\cos\theta_H<0.75$ and $\eta_{\rm
max}<2.5$, which corresponds to large rapidity gap events, contains a single
event.
  \label{lrg}
}
\end{figure}

\begin{figure}[p]
  \centerline{
\psfig{figure=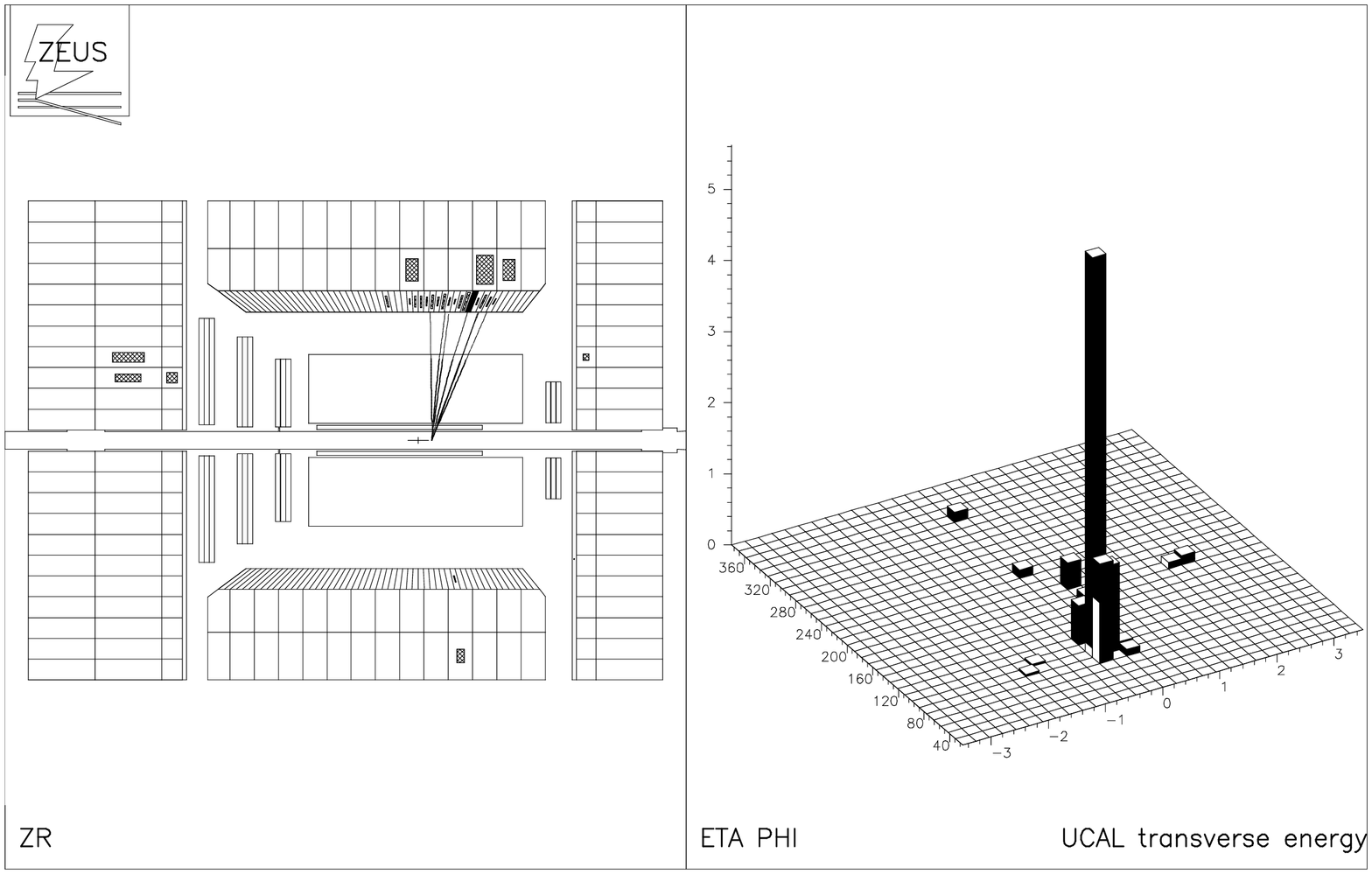,height=11.cm,angle=270}\kern-17.319cm
\hbox{\vrule height11.cm depth0.cm width0.5pt}\kern17.319cm\strut}
  \vskip20.mm
  \caption{
A display of the candidate large rapidity gap event. The left part shows the
ZEUS inner tracking system and the calorimeter. The proton beam enters from the
right; correspondingly, FCAL, BCAL and RCAL are displayed from left to right.
The filled rectangles in the calorimeter denote energy deposits which are above
the noise thresholds described in the text (cf.~section~3.1) and in addition are
above $100\,\mev$ if they are isolated.  The $ZR$--view shows a projection of
tracks and energy deposits along constant values of $Z$ and $R=(X^2+Y^2)^{1/2}$,
where the two hemispheres are separated by a plane perpendicular to the event
sphericity axis ($20^\circ<\phi<200^\circ$ corresponding to the upper half of
the display).  In the right part of the figure the $\eta$--$\phi$--distribution
of the calorimeter transverse energy is presented.  The FCAL cluster at
$\eta=2.2$ has $0.25\,\gev$ of transverse momentum. The kinematic variables for
this event are reconstructed to be $\ptmiss\eql14\pm 2\,\gev$,
$Q^2\eql300\pm70\,\gev^2$, $x\eql0.0093\pm0.0015$ and $y\eql0.35\pm0.10$.
  \label{lrgevt} 
}
\end{figure}

Figure \ref{lrg} shows a scatter plot of $\cos\theta_H$ versus $\eta_{\rm max}$
for the combined $e^+p$ and $e^-p$ samples.  The rectangle indicates the region
$\eta_{\rm max}\less2.5$ and $\cos\theta_H\less0.75$, which is taken from
\cite{ZEUSDiffSF} as the definition of LRG events (``LRG requirement'').  One
event of the $e^+p$ sample, shown in figure~\ref{lrgevt}, passes these LRG cuts.
The reconstructed kinematic variables for this event are $\ptmiss\eql14\pm
2\,\gev$, $Q^2\eql300\pm70\,\gev^2$, $x\eql0.0093\pm0.0015$ and $y\eql0.35\pm
0.10$.

In order to estimate event selection efficiencies and expected event rates in
the $e^+p$ case, the LRG analysis was also applied to the following CC MC sets
(at the generator level, $\eta_{\rm max}$ was defined as the maximum
pseudorapidity of all particles with momenta greater than $400\,\mev$ and with
$\eta<4.5$, where the latter requirement excluded the proton in diffractive
reactions):
\begin{itemize}
\item
{\sc ariadne}. This set uses {\sc lepto} for the differential cross section and
includes the simulation of diffractive--like processes via soft color
interactions between the perturbatively produced partons and color--charges in
the proton remnant \cite{SoftCol}.  The fraction of events with $\eta_{\rm
max}<2.5$ at the generator level was found to be $1.3\%$.
\item
{\sc meps}. This set also is based on {\sc lepto}, but does not simulate soft
color interactions. Here the rapidity gap between proton direction and hadronic
system is exponentially suppressed. At the generator level, $0.45\%$ of the
events had $\eta_{\rm max}<2.5$. Most of the events passing the LRG requirement
had a smaller rapidity gap at the generator level.
\item
{\sc pompyt}. This generator simulates only diffractive reactions and is based
on a factorizable model for high--energy diffractive processes where, within the
{\sc pythia} framework, the proton emits a pomeron, whose constituents take part
in a hard scattering process with the $W$ boson. For this analysis a hard quark
density distribution in the pomeron was assumed ($\propto\beta(1-\beta)$, where
$\beta$ is the momentum fraction of the quark relative to the pomeron momentum),
which in addition satisfied the momentum sum rule.\footnote{The latter
assumption conflicts with measurements \cite{ZEUSDiffSF} of the NC diffractive
cross section. Hence {\sc pompyt} predictions of absolute cross sections were
not used.}

It is interesting to note that, according to the {\sc pompyt} simulation, only
$18\%$ of all CC diffractive events with $Q^2>200\,\gev^2$ have $\eta_{\rm
max}<2.5$ at the generator level, $41\%$ of which pass the CC selection and
fulfill the LRG requirement.
\end{itemize}

The fractions of LRG events are summarized in table~\ref{tablrg} (for the $e^+p$
NC control sample this ratio was determined using the event weights as described
in section~3.3).  They roughly agree between the CC data, the {\sc ariadne}
simulation and the NC control sample, whereas for the {\sc meps} sample, the
fraction of LRG events is below that predicted in {\sc ariadne}.

The selection efficiency $\varepsilon_{\rm LRG}$, defined as
\begin{displaymath}
  \varepsilon_{\rm LRG}={\hbox{no.~of events with $Q^2_{\rm cor}>200\,\gev^2$
                         which pass CC cuts and LRG requirement}
                         \over
                         \hbox{no.~of events which have $Q^2>200\,\gev^2$ and 
                         $\eta_{\rm max}<2.5$ at generator level}}\;,
\end{displaymath}
is also shown in table~\ref{tablrg} for the three MC sets. Using the {\sc
ariadne} value of $\varepsilon_{\rm LRG}=0.43$, the cross section for $e^+p$ CC
interactions with $\eta_{\rm max}< 2.5$ was calculated to be
\begin{equation}
  \sigep(Q^2>200\,\gev^2;\eta_{\rm max}<2.5)=
  0.8\,{}_{-0.7}^{+1.8}\,{\rm[stat]}\,\pm0.1\,{\rm[syst]}\,{\rm pb}.
\end{equation}
The systematic uncertainty of this result was derived from the difference of the
efficiencies $\varepsilon_{\rm LRG}$ obtained using the {\sc meps} and {\sc
pompyt} sets.

\section{Summary and Conclusions}

The cross sections for the deep inelastic charged--current interactions
$e^+p\to\annu X$ and $e^-p\to\nu X$ have been measured in the kinematic region
$x>0.006$ and $Q^2>200\,\gev^2$, using the $e^+p$ and $e^-p$ data collected with
the ZEUS detector during 1993 and 1994. For the integrated cross sections,
\begin{eqnarray*}
 \sigep(Q^2\greater200\,\gev^2)&=&
 30.3\,{}^{\pls\chax5.5}_{\mns\chax4.2}\,{\rm [stat]}\;
           {}^{\pls1.6}_{\mns2.6}\,{\rm [syst]}\,{\rm pb}\\
 \sigem(Q^2\greater200\,\gev^2)&=&
 54.7\,{}^{\pls15.9}_{\mns\chax9.8}\,{\rm [stat]}\;
           {}^{\pls2.8}_{\mns3.4}\,{\rm [syst]}\,{\rm pb}\\
\end{eqnarray*}
are obtained, in good agreement with the Standard Model expectations of 
$32.3\,{\rm pb}$ and $65.8\,{\rm pb}$.

The differential cross sections have been measured as functions of $Q^2$, $x$
and $y$. The differences between $e^+p$ and $e^-p$ scattering, which are
expected from the $V\mns A$ helicity structure of the weak interaction and
from the quark content of the proton, are clearly observed. No significant
deviation from the Standard Model predictions is found.

The measured $Q^2$ dependence of the differential cross sections for $e^+$ and
$e^-$ induced CC DIS reactions is used to determine the $W$ mass. The measured
value of $M_W$ is $79\,{}^{+8}_{-7}\,{}^{+4}_{-4}\,\gev$, in good agreement
with direct measurements performed at $p\overline p$ colliders.

A jet analysis has been performed and CC events with multiple jets have been
observed. The distributions of the jet transverse momentum and rapidity as well
as transverse energy flow jet profiles have been measured and are compared to
the ${\cal O}(\alpha_s)$ predictions of the {\sc ariadne} and {\sc meps} Monte
Carlo models. Good agreement is observed in both the jet profiles and the jet
rates.

A search for charged--current events with a large rapidity gap between the
observed hadronic system and the outgoing proton system yielded one candidate
event in the $e^+p$ sample, corresponding to a cross section of $\sigep(Q^2>
200\,\gev^2;\eta_{\rm max}<2.5)=0.8\,{}_{-0.7}^{+1.8}\,{\rm[stat]}\,\pm0.1\,
{\rm[syst]}\,{\rm pb}$.

\vskip 2.cm
\leftline{\Large\bf Acknowledgments}
\vskip4.mm
\noindent
We appreciate the contributions to the construction and maintenance of the ZEUS
detector of the many people who are not listed as authors. The HERA machine
group and the DESY computing staffs are especially acknowledged for their
efforts to provide excellent operation of the collider and the data analysis
environment.  We thank the DESY directorate for strong support and
encouragement.


\vfill\eject
\begin{table}
  \begin{center}
  \begin{tabular} {||c||c|c||c|c||}
  \hline
  \stru{2.2ex}{1.2ex}cut        &\multicolumn{2}{c||}{$e^+p$}&
                                 \multicolumn{2}{c||}{$e^-p$}\\
  \cline{2-5}
  \stru{2.2ex}{1.2ex}description&data&MC($Q^2\greater200\,\gev^2$)&
                                 data&MC($Q^2\greater200\,\gev^2$)\\
  \hline\hline
    $Q^2>200\,\gev^2$ (MC only)                          
  &---&100\chapx\%&---&100\chapx\%\\\hline
    Trigger and preselection
  &   6307   &\chax83.2\%&   1450  &\chax86.1\%\\\hline
    $\ptmiss>11\,\gev$               
  &   2981   &\chax80.0\%&\cha940  &\chax83.1\%\\\hline
    $|\zvtx|<45\,{\rm cm}$             
  &   1283   &\chax71.0\%&\cha500  &\chax74.7\%\\\hline
    $\ptmiss^{\rm out}/\ptmiss>0.6$  
  &\cha958   &\chax70.3\%&\cha301  &\chax73.9\%\\\hline
    good track                       
  &\cha741   &\chax68.2\%&\cha237  &\chax71.7\%\\\hline
    $|\Delta\phi|<1$                 
  &\cha342   &\chax66.8\%&\cha113  &\chax70.3\%\\\hline
    $y_{JB}<0.8$                     
  &\cha279   &\chax66.2\%&\chaa98  &\chax69.2\%\\\hline
    $\ptmiss/E_t>0.4$                
  &\cha208   &\chax65.0\%&\chaa59  &\chax68.1\%\\\hline
    $P_t^{\rm tracks}/\ptmiss>0.1$  
  &\chaa87   &\chax63.0\%&\chaa36  &\chax66.3\%\\\hline
    Muon,timing,sparks               
  &\chaa59   &\chax63.0\%&\chaa30  &\chax66.3\%\\\hline
    Visual scan (data only)
  &\chaa56   &---                  &\chaa30  &---                  \\\hline
  \end{tabular}
\vskip1.cm
  \caption{
Data reduction by the selection cuts described in section~3.2.  The columns
denoted ``data'' indicate the numbers of events remaining after applying the cuts.
The ``MC'' columns show the  cumulative selection efficiencies for the {\sc ariadne} MC
simulation.
\label{cuteff}
}
  \end{center}
\end{table}
\begin{table}[p]
 \begin{center}
 \begin{tabular}{||c|c|c|c|c|c|c||}
 \hline
                      &\multicolumn{6}{c||}{\stru{2.2ex}{1.2ex}$Q^2\;[\gev^2]$}\\
 \cline{2-7}
         &$\chaxx200$&$\chaxx437$&$\chaxx955$&$\chax2089$&$\chax4517$&\\
  \noalign{\vskip-4.mm}
  {\Huge{$e^+p$}}       
         &\chax$-$   &\chax$-$   &\chax$-$   &$-$        &$-$        &$>200$\\          
  \noalign{\vskip-1.mm}
         &$\chaxx437$&$\chaxx955$&$\chax2089$&$\chax4517$&$    10000$&\\
 \hline\hline
 \stru{2.2ex}{1.2ex}$N_{\rm evt}$
   &$\chax8\chapx               $&$     15\chapx$
   &$    12\chapx               $&$     14\chapx$
   &$\chax7\chapx               $&$     56\chapx$\\        
 \hline
 \stru{2.2ex}{1.2ex}$\langle Q^2\rangle\;[\gev^2]$
   &$\cha310                    $&$\cha670    $
   &$   1450                    $&$   3070    $
   &$   6510                    $&$   1910    $\\        
 \hline
 \stru{2.2ex}{1.2ex}$N_{\rm bg}$
   &$\cha1.2                    $&$\cha0.2    $
   &$\cha0\chapx                $&$\cha0\chapx$
   &$\cha0\chapx                $&$\cha1.4    $\\        
 \hline
 \stru{2.2ex}{1.2ex}$\zeta$
   &$0.38                       $&$0.58    $
   &$0.70                       $&$0.76    $
   &$0.71                       $&$0.62    $\\        
 \hline
 \stru{2.2ex}{1.2ex}$\xi$
   &$0.76                       $&$0.72    $
   &$0.73                       $&$0.76    $
   &$0.79                       $&$1.0\chax$\\
\hline 
\hline
\stru{2.8ex}{1.8ex}$\sigma^{\rm exp}\,[{\rm pb}]$
   &$\chax6.1$&$\chax8.7$&$\chax5.8$
   &$\chax6.3$&$\chax3.4$&$    30.3$\\
 \hline
 \stru{2.8ex}{1.8ex}$\Delta\sigma^{\rm exp}\,{\rm[stat]}\;[{\rm pb}]$
   &${}^{\pls\chax3.0}_{\mns\chax2.2}$
   &${}^{\pls\chax2.9}_{\mns\chax2.3}$
   &${}^{\pls\chax2.2}_{\mns\chax1.7}$
   &${}^{\pls\chax2.2}_{\mns\chax1.7}$
   &${}^{\pls\chax1.8}_{\mns\chax1.3}$
   &${}^{\pls\chax5.5}_{\mns\chax4.2}$\\
 \hline
 \stru{2.8ex}{1.8ex}$\Delta\sigma^{\rm exp}\,{\rm[syst]}\;[{\rm pb}]$
   &${}^{\pls\chax1.3}_{\mns\chax2.7}$
   &${}^{\pls\chax0.6}_{\mns\chax0.6}$
   &${}^{\pls\chax0.3}_{\mns\chax0.3}$
   &${}^{\pls\chax0.6}_{\mns\chax0.3}$
   &${}^{\pls\chax0.7}_{\mns\chax0.4}$
   &${}^{\pls\chax1.6}_{\mns\chax2.6}$\\
 \hline
 \stru{2.2ex}{1.2ex}$\sigma^{\rm th}\,[{\rm pb}]$
   &$ \chax6.2                 $&$ \chax8.3$
   &$ \chax8.3                 $&$ \chax6.4$
   &$ \chax2.5                 $&$     32.3$\\
 \hline\hline
 \multicolumn{7}{||c||}{\stru{3.2ex}{2.2ex} \kern1.8cm Systematic checks}\\
 \hline
 \stru{2.8ex}{1.8ex} Check&\multicolumn{6}{c||}{$\Delta\sigma^{\rm exp}$\ [\%]}\\
 \hline\hline
 \stru{2.2ex}{1.2ex}$E$ scale $\pls3\%$ ($\pls5\%$ in FCAL)
   &$\chas12.6                 $&$\chasx6.3$
   &$\chasx2.1                 $&$-\chax3.3$
   &$-    10.8                 $&$\chasx2.1$\\
 \hline
 \stru{2.2ex}{1.2ex}$E$ scale $\mns3\%$ ($\mns5\%$ in FCAL)
   &$-    18.7                 $&$-\chax3.0$
   &$-\chax3.3                 $&$\chasx3.8$
   &$\chas13.6                 $&$-\chax2.0$\\
 \hline
 \stru{2.2ex}{1.2ex} Unfolding
   &$-\chax0.9                 $&$\chasx1.1$
   &$\chasx1.7                 $&$-\chax1.5$
   &$\chas13.8                 $&$-\chax1.0$\\
 \hline
 \stru{2.2ex}{1.2ex} $Q^2,x,y$ reconstruction
   &$-    15.7                 $&$-\chax2.1$
   &$\chasx2.0                 $&$\chasx7.6$
   &$\chasx5.8                 $&$-\chax0.3$\\
 \hline
 \stru{2.2ex}{1.2ex} $N_{\rm bg}+\Delta N_{\rm bg}$
   &$-    36.2                 $&$-\chax3.4$
   &$\chasx0.0                 $&$\chasx0.0$
   &$\chasx0.0                 $&$-\chax7.3$\\
 \hline
 \stru{2.2ex}{1.2ex} $N_{\rm bg}-\Delta N_{\rm bg}$
   &$\chas16.1                 $&$\chasx1.5$
   &$\chasx0.0                 $&$\chasx0.0$
   &$\chasx0.0                 $&$\chasx3.3$\\
 \hline
 \end{tabular}
\vskip1.cm
\caption{
Cross sections for CC $e^+p$ scattering in intervals of $Q^2$, and for
$Q^2>200\,\gev^2$. The rows denoted $N_{\rm evt}$, $\langle Q^2\rangle$, $N_{\rm
bg}$, $\zeta$, $\xi$, $\sigma^{\rm exp}$, and $\sigma^{\rm th}$ show the raw
event numbers, the mean $Q^2$ according to the SM prediction, the estimated
number of background events from photoproduction, the bin--by--bin acceptance
corrections, the purities, the experimental cross sections, and the theoretical
cross sections, respectively. For $\sigma^{\rm exp}$ the statistical errors
$\Delta\sigma^{\rm exp}\,{\rm[stat]}$ and systematic uncertainties
$\Delta\sigma^{\rm exp}\,{\rm[syst]}$ are quoted. The latter are calculated from
the deviations resulting from the systematic studies described in the text
(including the luminosity uncertainty of $3\%$ and the overall uncertainty of
$3\%$ assigned to the efficiency of those selection cuts involving track
quantities). The dominant contributions are detailed in the lower part of the
table.
\label{resQtotpos}
}
\end{center}
\end{table}
\begin{table}[p]
 \begin{center}
 \begin{tabular}{||c|c|c|c|c|c|c|c||}
 \hline
                      &\multicolumn{7}{c||}{\stru{2.2ex}{1.2ex}$Q^2\;[\gev^2]$}\\
 \cline{2-8}
         &$\chaxx200$&$\chaxx437$&$\chaxx955$&$\chax2089$&$\chax4517$&   $   10000$&\\
  \noalign{\vskip-4.mm}
  {\Huge{$e^-p$}}       
         &\chax$-$   &\chax$-$   &\chax$-$   &$-$        &$-$      &$-$ &$>200$\\
  \noalign{\vskip-1.mm}
         &$\chaxx437$&$\chaxx955$&$\chax2089$&$\chax4517$&$    10000$&   $    21877$&\\
 \hline\hline
 \stru{2.2ex}{1.2ex}$N_{\rm evt}$
   &$\chax3\chapx                     $&$\chax3\chapx$
   &$\chax7\chapx                     $&$\chax6\chapx$
   &$\chax9\chapx                     $&$\chax2\chapx$  &$      30\chapx$\\        
 \hline
 \stru{2.2ex}{1.2ex}$\langle Q^2\rangle\;[\gev^2]$
   &$\chaxx310                        $&$\chaxx670    $
   &$\chax1470                        $&$\chax3140    $
   &$\chax6700                        $&$    13900    $&$\chax3340    $\\        
 \hline
 \stru{2.2ex}{1.2ex}$N_{\rm bg}$
   &$\cha0.3                          $&$\cha0.1     $
   &$\cha0\chapx                      $&$\cha0\chapx $
   &$\cha0\chapx                      $&$\cha0\chapx $  &$\cha0.4$\\        
 \hline
 \stru{2.2ex}{1.2ex}$\zeta$
   &$0.39                       $&$0.56  $
   &$0.74                       $&$0.78  $
   &$0.74                       $&$0.59  $  &$0.66    $\\        
 \hline
 \stru{2.2ex}{1.2ex}$\xi$
   &$0.74                       $&$0.73  $
   &$0.70                       $&$0.74  $
   &$0.80                       $&$0.83  $  &$1.0\chax$\\
 \hline
 \stru{2.8ex}{1.8ex}$\sigma^{\rm exp}\,[{\rm pb}]$
   &$\chax8.4$&$\chax6.3$&$    11.6$
   &$\chax9.4$&$    14.9$&$\chax4.1$
   &$    54.7$\\
 \hline
 \stru{2.8ex}{1.8ex}$\Delta\sigma^{\rm exp}\,{\rm[stat]}\;[{\rm pb}]$
   &${}^{\pls\chax8.2}_{\mns\chax4.6}$
   &${}^{\pls\chax6.2}_{\mns\chax3.4}$
   &${}^{\pls\chax6.3}_{\mns\chax4.3}$
   &${}^{\pls\chax5.6}_{\mns\chax3.7}$
   &${}^{\pls\chax6.8}_{\mns\chax4.9}$
   &${}^{\pls\chax5.4}_{\mns\chax2.7}$
   &${}^{\pls    15.9}_{\mns\chax9.8}$\\
 \hline
 \stru{2.8ex}{1.8ex}$\Delta\sigma^{\rm exp}\,{\rm[syst]}\;[{\rm pb}]$
   &${}^{\pls\chax1.6}_{\mns\chax2.7}$
   &${}^{\pls\chax0.6}_{\mns\chax0.7}$
   &${}^{\pls\chax0.7}_{\mns\chax0.7}$
   &${}^{\pls\chax0.8}_{\mns\chax0.6}$
   &${}^{\pls\chax1.9}_{\mns\chax1.6}$
   &${}^{\pls\chax1.5}_{\mns\chax1.2}$
   &${}^{\pls\chax2.8}_{\mns\chax2.9}$\\
 \hline
 \stru{2.2ex}{1.2ex}$\sigma^{\rm th}\,[{\rm pb}]$
   &$\chax7.9                   $&$    12.2$
   &$    15.1                   $&$    15.7$
   &$    10.6                   $&$\chax3.7$   &$    65.8$\\
 \hline\hline
 \multicolumn{8}{||c||}{\stru{3.2ex}{2.2ex} \kern1.8cm Systematic checks}\\
 \hline
 \stru{2.8ex}{1.8ex} Check&\multicolumn{7}{c||}{$\Delta\sigma^{\rm exp}$\ [\%]}\\
 \hline\hline
 \stru{2.2ex}{1.2ex}$E$ scale $\pls3\%$ ($\pls5\%$ in FCAL)
   &$\chas14.2                 $&$\chasx7.1$
   &$\chasx2.8                 $&$\chasx0.9$
   &$-\chax3.9                 $&$-    24.9$   &$\chasx2.0$\\
 \hline
 \stru{2.2ex}{1.2ex}$E$ scale $\mns3\%$ ($\mns5\%$ in FCAL)
   &$-    10.9                 $&$-\chax7.8$
   &$-\chax2.6                 $&$\chasx1.2$
   &$\chasx9.5                 $&$\chas26.9$   &$\chasx1.1$\\
 \hline
 \stru{2.2ex}{1.2ex} Unfolding
   &$-\chax0.5                 $&$\chasx0.0$
   &$-\chax3.3                 $&$-\chax4.2$
   &$-\chax9.3                 $&$\chas24.3$   &$-\chax2.1$\\
 \hline
 \stru{2.2ex}{1.2ex} $Q^2,x,y$ reconstruction
   &$-    15.2                 $&$-\chax5.9$
   &$\chasx2.2                 $&$\chasx6.8$
   &$\chasx7.8                 $&$-    14.5$   &$-\chax1.0$\\
 \hline
 \stru{2.2ex}{1.2ex} $N_{\rm bg}+\Delta N_{\rm bg}$
   &$-    25.9                 $&$-\chax5.0$
   &$\chasx0.0                 $&$\chasx0.0$
   &$\chasx0.0                 $&$\chasx0.0$   &$-\chax4.0$\\
 \hline
 \stru{2.2ex}{1.2ex} $N_{\rm bg}-\Delta N_{\rm bg}$
   &$\chas11.1                 $&$\chasx2.7$
   &$\chasx0.0                 $&$\chasx0.0$
   &$\chasx0.0                 $&$\chasx0.0$   &$\chasx1.7$\\
 \hline
 \end{tabular}
\vskip1.cm
\caption{
Cross sections for CC $e^-p$ scattering in intervals of $Q^2$, and for
$Q^2>200\,\gev^2$. The rows denoted $N_{\rm evt}$, $\langle Q^2\rangle$, $N_{\rm
bg}$, $\zeta$, $\xi$, $\sigma^{\rm exp}$, and $\sigma^{\rm th}$ show the raw
event numbers, the mean $Q^2$ according to the SM prediction, the estimated
number of background events from photoproduction, the bin--by--bin acceptance
corrections, the purities, the experimental cross sections, and the theoretical
cross sections, respectively. For $\sigma^{\rm exp}$ the statistical errors
$\Delta\sigma^{\rm exp}\,{\rm[stat]}$ and systematic uncertainties
$\Delta\sigma^{\rm exp}\,{\rm[syst]}$ are quoted. The latter are calculated from
the deviations resulting from the systematic studies described in the text
(including the luminosity uncertainty of $2\%$ and the overall uncertainty of
$3\%$ assigned to the efficiency of those selection cuts involving track
quantities). The dominant contributions are detailed in the lower part of the
table.
\label{resQtotele}
}
\end{center}
\end{table}
\begin{table}
 \begin{center}
 \begin{tabular}{||c|c|c|c|c|c|c||}
 \hline
                      &\multicolumn{6}{c||}{\stru{2.2ex}{1.2ex}$x$}\\
 \cline{2-7}
         &$0.006$&$0.014$ &$0.033$ &$0.076    $&$0.17$&$0.40$\\
  \noalign{\vskip-4.mm}
  {\Huge{$e^+p$}}       
         &$-$    &\chax$-$&\chax$-$&$-$        &$-$   &$-$ \\
  \noalign{\vskip-1.mm}
         &$0.014$&$0.033$ &$0.076$ &$0.17\chax$&$0.40$&$1.0\chax$\\
 \hline\hline
 \stru{2.2ex}{1.2ex}$N_{\rm evt}$
   &$\chax3\chapx                     $&$    20\chapx $
   &$    11\chapx                     $&$    13\chapx $
   &$\chax9\chapx                     $&$\chax0\chapx $\\        
 \hline
 \stru{2.2ex}{1.2ex}$\langle x\rangle$
   &$0.010                        $&$0.023    $
   &$0.052                        $&$0.115    $
   &$0.249                        $&$0.480    $\\
 \hline
 \stru{2.2ex}{1.2ex}$N_{\rm bg}$
   &$\cha1.4                          $&$\cha0\chapx$
   &$\cha0\chapx                      $&$\cha0\chapx$
   &$\cha0\chapx                      $&$\cha0\chapx$\\        
 \hline
 \stru{2.2ex}{1.2ex}$\zeta$
   &$0.41                       $&$0.78   $
   &$0.79                       $&$0.68   $
   &$0.48                       $&$0.18$\\        
 \hline
 \stru{2.2ex}{1.2ex}$\xi$
   &$0.68                       $&$0.72   $
   &$0.77                       $&$0.82   $
   &$0.88                       $&$0.84$\\
 \hline
 \stru{2.8ex}{1.8ex}$\sigma^{\rm exp}\,[{\rm pb}]$
   &$\chax1.3$&$\chax8.7$&$\chax4.7$&$\chax6.5$&$\chax6.4$&$\chax0.0$\\
 \hline
 \stru{2.8ex}{1.8ex}$\Delta\sigma^{\rm exp}\,{\rm[stat]}\;[{\rm pb}]$
   &${}^{\pls\chax1.3}_{\mns\chax0.7}$
   &${}^{\pls\chax2.4}_{\mns\chax1.9}$
   &${}^{\pls\chax1.9}_{\mns\chax1.4}$
   &${}^{\pls\chax2.3}_{\mns\chax1.8}$
   &${}^{\pls\chax2.9}_{\mns\chax2.1}$
   &${}^{\pls\chax3.5}_{\mns\chax0.0}$\\
 \hline
 \stru{2.8ex}{1.8ex}$\Delta\sigma^{\rm exp}\,{\rm[syst]}\;[{\rm pb}]$
   &${}^{\pls\chax1.0}_{\mns\chax1.3}$
   &${}^{\pls\chax0.5}_{\mns\chax0.7}$
   &${}^{\pls\chax0.4}_{\mns\chax0.2}$
   &${}^{\pls\chax0.5}_{\mns\chax0.7}$
   &${}^{\pls\chax0.8}_{\mns\chax0.6}$
   &${}^{\pls\chax0.0}_{\mns\chax0.0}$\\
 \hline
 \stru{2.2ex}{1.2ex}$\sigma^{\rm th}\,[{\rm pb}]$
   &$\chax3.4                   $&$\chax6.3$
   &$\chax8.4                   $&$\chax8.3$
   &$\chax4.2                   $&$\chax0.5$\\
 \hline
 \end{tabular}
\vskip8.mm
 \begin{tabular}{||c|c|c|c|c|c|c||}
 \hline
                      &\multicolumn{6}{c||}{\stru{2.2ex}{1.2ex}$x$}\\
 \cline{2-7}
         &$0.006$&$0.014$ &$0.033$ &$0.076    $&$0.17$&$0.40$\\
  \noalign{\vskip-4.mm}
  {\Huge{$e^-p$}}       
         &$-$    &\chax$-$&\chax$-$&$-$        &$-$   &$-$ \\
  \noalign{\vskip-1.mm}
         &$0.014$&$0.033$ &$0.076$ &$0.17\chax$&$0.40$&$1.0\chax$\\
 \hline\hline
 \stru{2.2ex}{1.2ex}$N_{\rm evt}$
   &$\chax3\chapx                     $&$\chax5\chapx $
   &$\chax5\chapx                     $&$     9\chapx $
   &$\chax8\chapx                     $&$\chax0\chapx $\\        
 \hline
 \stru{2.2ex}{1.2ex}$\langle x\rangle$
   &$0.010                        $&$0.024    $
   &$0.053                        $&$0.119    $
   &$0.256                        $&$0.489    $\\
 \hline
 \stru{2.2ex}{1.2ex}$N_{\rm bg}$
   &$\cha0.4                          $&$\cha0\chapx$
   &$\cha0\chapx                      $&$\cha0\chapx$
   &$\cha0\chapx                      $&$\cha0\chapx$\\        
 \hline
 \stru{2.2ex}{1.2ex}$\zeta$
   &$0.41                       $&$0.77   $
   &$0.84                       $&$0.72   $
   &$0.55                       $&$0.17   $\\        
 \hline
 \stru{2.2ex}{1.2ex}$\xi$
   &$0.61                       $&$0.73   $
   &$0.73                       $&$0.85   $
   &$0.87                       $&$0.83   $\\
 \hline
 \stru{2.8ex}{1.8ex}$\sigma^{\rm exp}\,[{\rm pb}]$
   &$\chax7.8$&$\chax7.9$&$\chax7.3$&$    15.3$&$    17.6$&$\chax0.0$\\
 \hline
 \stru{2.8ex}{1.8ex}$\Delta\sigma^{\rm exp}\,{\rm[stat]}\;[{\rm pb}]$
   &${}^{\pls\chax7.6}_{\mns\chax4.2}$
   &${}^{\pls\chax5.3}_{\mns\chax3.4}$
   &${}^{\pls\chax4.9}_{\mns\chax3.2}$
   &${}^{\pls\chax7.0}_{\mns\chax5.1}$
   &${}^{\pls\chax8.7}_{\mns\chax6.3}$
   &${}^{\pls    13.2}_{\mns\chax0.0}$\\
 \hline
 \stru{2.8ex}{1.8ex}$\Delta\sigma^{\rm exp}\,{\rm[syst]}\;[{\rm pb}]$
   &${}^{\pls\chax1.5}_{\mns\chax2.5}$
   &${}^{\pls\chax0.7}_{\mns\chax0.7}$
   &${}^{\pls\chax0.4}_{\mns\chax0.5}$
   &${}^{\pls\chax1.1}_{\mns\chax2.0}$
   &${}^{\pls\chax3.0}_{\mns\chax1.6}$
   &${}^{\pls\chax0.0}_{\mns\chax0.0}$\\
 \hline
 \stru{2.2ex}{1.2ex}$\sigma^{\rm th}\,[{\rm pb}]$
   &$\chax4.2                   $&$\chax8.4$
   &$    14.9                   $&$    19.2$
   &$    14.8                   $&$\chax2.7$\\
 \hline \end{tabular}
\vskip1.cm
\caption{
Cross sections for CC $e^+p$ (top) and $e^-p$ (bottom) scattering for
$Q^2>200\,\gev^2$ in intervals of $x$. The rows denoted $N_{\rm evt}$, $\langle
x\rangle$, $N_{\rm bg}$, $\zeta$, $\xi$, $\sigma^{\rm exp}$, and $\sigma^{\rm
th}$ show the raw event numbers, the mean $x$ according to the SM prediction,
the estimated number of background events from photoproduction, the bin--by--bin
acceptance corrections, the purities, the experimental cross sections, and the
theoretical cross sections, respectively. For $\sigma^{\rm exp}$ the statistical
errors $\Delta\sigma^{\rm exp}\,{\rm[stat]}$ and systematic uncertainties
$\Delta\sigma^{\rm exp}\,{\rm[syst]}$ are quoted.
\label{resx}
}
\end{center}
\end{table}
\begin{table}[p]
 \begin{center}
 \begin{tabular}{||c|c|c|c|c|c||}
 \hline
                      &\multicolumn{5}{c||}{\stru{2.2ex}{1.2ex}$y$}\\
 \cline{2-6}
         &$0.0$&$0.2$&$0.4$&$0.6$&$0.8$\\
  \noalign{\vskip-4.mm}
  {\Huge{$e^+p$}}       
         &$-\chax$&$-\chax$&$-\chax$&$-\chax$&$-\chax$\\
  \noalign{\vskip-1.mm}
         &$0.2$&$0.4$&$0.6$&$0.8$&$1.0$\\
 \hline\hline
 \stru{2.2ex}{1.2ex}$N_{\rm evt}$
   &$    14\chapx                     $&$    27\chapx $
   &$    11\chapx                     $&$\chax3\chapx $
   &$\chax1\chapx                     $\\        
 \hline
 \stru{2.2ex}{1.2ex}$\langle y\rangle$
   &$0.101                        $&$0.293    $
   &$0.496                        $&$0.695    $
   &$0.895                        $\\
 \hline
 \stru{2.2ex}{1.2ex}$N_{\rm bg}$
   &$\cha0\chapx                          $&$\cha0.2    $
   &$\cha1.2                              $&$\cha0\chapx$
   &$\cha0\chapx                      $\\        
 \hline
 \stru{2.2ex}{1.2ex}$\zeta$
   &$0.49                       $&$0.86  $
   &$0.79                       $&$0.73  $
   &$0.15                       $\\        
 \hline
 \stru{2.2ex}{1.2ex}$\xi$
   &$0.90                       $&$0.79  $
   &$0.77                       $&$0.70  $
   &$0.80                       $\\
 \hline
 \stru{2.8ex}{1.8ex}$\sigma^{\rm exp}\,[{\rm pb}]$
   &$    9.6$&$     10.7$&$\chax4.2$&$\chax1.4$&$\chax2.3$\\
 \hline
 \stru{2.8ex}{1.8ex}$\Delta\sigma^{\rm exp}\,{\rm[stat]}\;[{\rm pb}]$
   &${}^{\pls\chax3.3}_{\mns\chax2.5}$
   &${}^{\pls\chax2.5}_{\mns\chax2.1}$
   &${}^{\pls\chax1.7}_{\mns\chax1.3}$
   &${}^{\pls\chax1.4}_{\mns\chax0.8}$
   &${}^{\pls\chax5.4}_{\mns\chax1.9}$\\
 \hline
 \stru{2.8ex}{1.8ex}$\Delta\sigma^{\rm exp}\,{\rm[syst]}\;[{\rm pb}]$
   &${}^{\pls\chax0.5}_{\mns\chax0.7}$
   &${}^{\pls\chax0.6}_{\mns\chax0.5}$
   &${}^{\pls\chax0.6}_{\mns\chax1.0}$
   &${}^{\pls\chax0.2}_{\mns\chax0.1}$
   &${}^{\pls\chax0.6}_{\mns\chax1.1}$\\
 \hline
 \stru{2.2ex}{1.2ex}$\sigma^{\rm th}\,[{\rm pb}]$
   &$     11.8                   $&$ \chax8.3$
   &$ \chax5.4                   $&$ \chax3.7$
   &$ \chax3.0                   $\\
 \hline
 \end{tabular}
\vskip8.mm
 \begin{tabular}{||c|c|c|c|c|c||}
 \hline\hline
                      &\multicolumn{5}{c||}{\stru{2.2ex}{1.2ex}$y$}\\
 \cline{2-6}
         &$0.0$&$0.2$&$0.4$&$0.6$&$0.8$\\
  \noalign{\vskip-4.mm}
  {\Huge{$e^-p$}}       
         &$-\chax$&$-\chax$&$-\chax$&$-\chax$&$-\chax$\\
  \noalign{\vskip-1.mm}
         &$0.2$&$0.4$&$0.6$&$0.8$&$1.0$\\
 \hline\hline
 \stru{2.2ex}{1.2ex}$N_{\rm evt}$
   &$\chax6\chapx                     $&$\chax5\chapx $
   &$    11\chapx                     $&$\chax5\chapx $
   &$\chax3\chapx                     $\\        
 \hline
 \stru{2.2ex}{1.2ex}$\langle y\rangle$
   &$0.101                        $&$0.293    $
   &$0.496                        $&$0.696    $
   &$0.895                        $\\
 \hline
 \stru{2.2ex}{1.2ex}$N_{\rm bg}$
   &$\cha0\chapx                      $&$\cha0.1    $
   &$\cha0.3                          $&$\cha0\chapx$
   &$\cha0\chapx                      $\\        
 \hline
 \stru{2.2ex}{1.2ex}$\zeta$
   &$0.51                       $&$0.91   $
   &$0.81                       $&$0.77   $
   &$0.20                       $\\        
 \hline
 \stru{2.2ex}{1.2ex}$\xi$
   &$0.92                       $&$0.79   $
   &$0.81                       $&$0.74   $
   &$0.90                       $\\
 \hline
 \stru{2.8ex}{1.8ex}$\sigma^{\rm exp}\,[{\rm pb}]$
   &$    14.4$&$\chax6.6$&$    16.2$&$\chax7.9$&$\chax19.0$\\
 \hline
 \stru{2.8ex}{1.8ex}$\Delta\sigma^{\rm exp}\,{\rm[stat]}\;[{\rm pb}]$
   &${}^{\pls\chax8.6}_{\mns\chax5.7}$
   &${}^{\pls\chax4.4}_{\mns\chax2.9}$
   &${}^{\pls\chax6.4}_{\mns\chax4.9}$
   &${}^{\pls\chax5.2}_{\mns\chax3.4}$
   &${}^{\pls    18.5}_{\mns    10.3}$\\
 \hline
 \stru{2.8ex}{1.8ex}$\Delta\sigma^{\rm exp}\,{\rm[syst]}\;[{\rm pb}]$
   &${}^{\pls\chax1.2}_{\mns\chax1.6}$
   &${}^{\pls\chax0.3}_{\mns\chax0.4}$
   &${}^{\pls\chax1.3}_{\mns\chax1.4}$
   &${}^{\pls\chax0.8}_{\mns\chax0.5}$
   &${}^{\pls\chax4.4}_{\mns\chax7.6}$\\
 \hline
 \stru{2.2ex}{1.2ex}$\sigma^{\rm th}\,[{\rm pb}]$
   &$     23.7                   $&$     15.6$
   &$     11.0                   $&$ \chax8.7$
   &$ \chax6.9                   $\\
 \hline
 \end{tabular}
\vskip1.cm
\caption{
Cross sections for CC $e^+p$ (top) and $e^-p$ (bottom) scattering for
$Q^2>200\,\gev^2$ in intervals of $y$. The rows denoted $N_{\rm evt}$, $\langle
y\rangle$, $N_{\rm bg}$, $\zeta$, $\xi$, $\sigma^{\rm exp}$, and $\sigma^{\rm
th}$ show the raw event numbers, the mean $y$ according to the SM prediction,
the estimated number of background events from photoproduction, the bin--by--bin
acceptance corrections, the purities, the experimental cross sections, and the
theoretical cross sections, respectively. For $\sigma^{\rm exp}$ the statistical
errors $\Delta\sigma^{\rm exp}\,{\rm[stat]}$ and systematic uncertainties
$\Delta\sigma^{\rm exp}\,{\rm[syst]}$ are quoted.
\label{resy}
}
\end{center}
\end{table}
\begin{table}[p]
\begin{center}
\begin{tabular}{||c||c|c|c||}
  \hline
  \stru{3.ex}{2.ex}Distribution&\hbox to 8.mm{\hfil$\chi^2$\hfil}&$N_{\rm DOF}$&
                   $p(\chi^2,N_{\rm DOF})$\stru{2.5ex}{2.ex}\\
  \hline\hline
  $d\sigma/d\,Q^2\;(e^+p)$&1.9& 5&0.86\stru{2.5ex}{2.ex}\\
  $d\sigma/d\,Q^2\;(e^-p)$&3.2& 6&0.78\stru{2.5ex}{2.ex}\\
  \hline                         
  $d\sigma/d\,x  \;(e^+p)$&9.9& 5&0.08\stru{2.5ex}{2.ex}\\
  $d\sigma/d\,x  \;(e^-p)$&3.7& 5&0.59\stru{2.5ex}{2.ex}\\
  \hline                         
  $d\sigma/d\,y  \;(e^+p)$&5.0& 5&0.42\stru{2.5ex}{2.ex}\\
  $d\sigma/d\,y  \;(e^-p)$&7.8& 5&0.17\stru{2.5ex}{2.ex}\\
  \hline
\end{tabular}
\caption{
Quality of agreement between experimental results and SM predictions: for the
various differential cross sections $\chi^2$, $N_{\rm DOF}$, and $p(\chi^2,
N_{\rm DOF})$ denote the $\chi^2$ value calculated using statistical errors
only, the number of degrees of freedom, and the corresponding
$\chi^2$--probability.
\label{tabchi}
}
\end{center}
\end{table}
\vfill\eject
\begin{table}[p]
\begin{center}
\centerline{\vbox{\hsize17.cm
\begin{tabular}{||c||c|c|c|c||c|c|c|c||}
  \hline
  \stru{3.ex}{2.ex}&\multicolumn{4}{c||}{$e^+p$}&\multicolumn{4}{c||}{$e^-p$}\\
  \hline
  \stru{2.6ex}{1.8ex}$N_{\rm jets}$
   &$N_{\rm evt}$&$R_{\rm jet}^{\rm data}\;[\%]$&
   \multicolumn{2}{c||}{$R_{\rm jet}^{\rm MC}\;[\%]$}
   &$N_{\rm evt}$&$R_{\rm jet}^{\rm data}\;[\%]$&
   \multicolumn{2}{c||}{$R_{\rm jet}^{\rm MC}\;[\%]$}\\
   &&&{\sc aria.}&{\sc meps}&&&{\sc aria.}&{\sc meps}\\
  \hline\hline
  \stru{2.6ex}{1.8ex}
  0 &$\chax0$&$\chax0.0\,{}^{+3.2}_{-0.0}$&$\chax0.4\pm0.1$&$\chax0.4\pm0.1$
    &$\chax0$&$\chax0.0\,{}^{+6.0}_{-0.0}$&$\chax0.3\pm0.1$&$\chax0.4\pm0.1$\\
  \hline
  \stru{2.6ex}{1.8ex}
  1 &$    49$&$    87.5\,{}^{+4.5}_{-6.1}$&$    90.9\pm0.4$&$    82.2\pm0.6$
    &$    27$&$    90.0\,{}^{+5.4}_{-8.8}$&$    91.6\pm0.4$&$    83.3\pm0.5$\\
  \hline
  \stru{2.6ex}{1.8ex}
  2 &$\chax6$&$    10.7\,{}^{+5.9}_{-4.2}$&$\chax8.4\pm0.4$&$    16.3\pm0.5$
    &$\chax3$&$    10.0\,{}^{+8.8}_{-5.4}$&$\chax7.8\pm0.4$&$    14.6\pm0.5$\\
  \hline
  \stru{2.6ex}{1.8ex}
  3 &$\chax1$&$\chax1.8\,{}^{+4.0}_{-1.5}$&$\chax0.3\pm0.1$&$\chax1.1\pm0.2$
    &$\chax0$&$\chax0.0\,{}^{+6.0}_{-0.0}$&$\chax0.3\pm0.1$&$\chax1.2\pm0.2$\\
  \hline
\end{tabular}
}}
\caption{
Jet rates for $e^+p$ and $e^-p$ data.  The columns denoted $N_{\rm evt}$ show
the numbers of events in the CC samples which have $N_{\rm jets}$ jets in
addition to the proton remnant (for the jet definition see text). The $R_{\rm
jet}^{\rm data}$ are the corresponding jet rates, which are also given for the
{\sc ariadne} and {\sc meps} MC simulations ($R_{\rm jet}^{\rm MC}$). Only
statistical errors are shown.
\label{jetst}
}
\end{center}
\vskip30.mm
\begin{center}
\begin{tabular}{||l||l|c||}
  \hline
  \multicolumn{1}{||c||}{\stru{3.ex}{2.ex}data set}
   &\hbox to 2.2cm{\hfil$R_{\rm LRG}\,[\%]$\hfil}
   &\hbox to 2.2cm{\hfil$\varepsilon_{\rm LRG}\,[\%]$\hfil}\\
  \hline\hline
  \stru{2.6ex}{1.8ex}CC $e^+p$
    &$\chax1.8\,{}^{+\;4.0}_{-\;1.5}$&$\none$\\
  \hline
  \stru{2.6ex}{1.8ex}{\sc ariadne} $e^+p$
    &$\chax0.9\,{}^{+\;0.2}_{-\;0.1}$&$43$\\
  \hline
  \stru{2.6ex}{1.8ex}{\sc meps} $e^+p$
    &$\chax0.3\pm0.1                $&$39$\\
  \hline
  \stru{2.6ex}{1.8ex}{\sc pompyt} $e^+p$
    &$    13.8\pm0.5                $&$53$\\
  \hline
  \stru{2.6ex}{1.8ex}NC control sample ($e^+p$)
    &$\chax0.9\pm0.4                $&$\none$\\
  \hline
\end{tabular}
\caption{
The fraction of large rapidity gap (LRG) events in the $e^+p$ CC sample, in the
{\sc ariadne}, {\sc meps} and {\sc pompyt} CC MC sets, and in the NC control
sample. $R_{\rm LRG}$ denotes the fraction of events with $\eta_{\rm max}<2.5$
and $\cos\theta_H<0.75$ (LRG events) in the total samples passing the CC
selection cuts. For the NC control sample, $R_{\rm LRG}$ is calculated from the
sums of cross--section weights as described in section~3.3. The selection
efficiency $\varepsilon_{\rm LRG}$ is defined as the ratio of the number of
events which pass the CC cuts, fulfil the LRG requirement and have $Q^2_{\rm
cor}>200\,\gev^2$ to the number of events having $\eta_{\rm max}<2.5$ and
$Q^2>200\,\gev^2$ at the generator level.
\label{tablrg}
}
\end{center}
\end{table}
\vfill\eject

\newpage
\begin{thebibliography}{99}

\bibitem{FTneut}
CDHS Collaboration, H.Abramowicz et al., Z.~Phys.~{\bf C25}(1984)29;\\
CDHSW Collaboration, J.P.Berge et al., Z.~Phys.~{\bf C49}(1991)187;\\
CCFR Collaboration, E.Oltman et al., Z.~Phys.~{\bf C53}(1992)51;\\
BEBC WA21 Collaboration, G.T.Jones et al., Z.~Phys.~{\bf C62}(1994)575.

\bibitem{H1CC93}
H1 Collaboration, T.Ahmed et al., Phys.~Lett.~{\bf B}324(1994)241.

\bibitem{ZEUSCC93}
ZEUS Collaboration, M.Derrick et al., Phys.~Rev.~Lett.~{\bf 75}(1995)1006.

\bibitem{H1CC94}
H1 Collaboration, S.Aid et al., Z.~Phys.~{\bf C67}(1995)565.

\bibitem{H1CC94Q}
H1 Collaboration, S.Aid et al., DESY 96--046 (1996).

\bibitem{mwppbar} CDF Collaboration, F.Abe et al., Phys.~Rev.~Lett.~{\bf 75}(1995)11;\\
                  CDF Collaboration, F.Abe et al., Phys.~Rev.~{\bf D52}(1995)4784;\\
                  D0 Collaboration, S.Abachi et al., Proc.~Topical Workshop on 
                     Proton--Antiproton Collider Physics, Tsukuba, Japan (1993);\\
                  UA2 Collaboration, J.Alitti et al., Phys.~Lett~{\bf B241}(1990)150.

\bibitem{lrgdis} ZEUS Collaboration, M.Derrick et al., Phys.~Lett.~{\bf B315}(1993)481;\\
                 ZEUS Collaboration, M.Derrick et al., Phys.~Lett.~{\bf B332}(1994)228;\\
                 H1   Collaboration, T.Ahmed et al., Nucl.~Phys.~{\bf B429}(1994)477.

\bibitem{lrgquarks}  ZEUS Collaboration, M.Derrick et al., Phys.~Lett.~{\bf B346}(1995)399;\\
                     H1   Collaboration, T.Ahmed et al., Phys.~Lett.~{\bf B348}(1995)681;\\
                     ZEUS Collaboration, M.Derrick et al., Phys.~Lett.~{\bf B356}(1995)129.

\bibitem{ZEUSDiffSF} ZEUS Collaboration, M.Derrick et al., Z.~Phys.~{\bf C68}(1995)569.

\bibitem{ZEUSDiffXS} ZEUS Collaboration, M.Derrick et al., DESY 96--018 (1996).

\bibitem{ZEUSF293} ZEUS Collaboration, M.Derrick et al., Z.~Phys.~{\bf C65}(1995)379.

\bibitem{ZEUS} ZEUS Collaboration, M.Derrick et al., Phys.~Lett.~{\bf B293}(1992)465.

\bibitem{trigger} The ZEUS Detector, Status Report 1993, DESY (1993).

\bibitem{CAL} M.Derrick et al., Nucl.~Inst.~Meth.~{\bf A309}(1991)77;\\
              A.Andresen et al., Nucl.~Inst.~Meth.~{\bf A309}(1991)101;\\
              A.Bernstein et al., Nucl.~Inst.~Meth.~{\bf A336}(1993)23;\\
              A.Caldwell et al., Nucl.~Inst.~Meth.~{\bf A321}(1992)356.

\bibitem{BAC} H.Abramowicz et al., Nucl.~Inst.~Meth.~{\bf A313}(1992)126.

\bibitem{VXD} C.Alvisi et al., Nucl.~Inst.~Meth.~{\bf A305}(1991)30.

\bibitem{CTD} N.Harnew et al., Nucl.~Inst.~Meth.~{\bf A279}(1989)290;\\
              B.Foster et al., Nucl.~Phys.,~Proc.~Suppl.~{\bf B32}(1993);\\
              B.Foster et al., Nucl.~Inst.~Meth.~{\bf A338}(1994)254.

\bibitem{LUMI} J.Andruszk\'{o}w, DESY 92--066 (1992).

\bibitem{lepto} {\sc lepto 6.3}: G.Ingelman, 
Proc.~1991 Workshop on Physics at HERA, ed. W.Buchm\"uller and G.Ingelman
(DESY, Hamburg, 1992), Vol.~3, p.1366.

\bibitem{heracles} {\sc heracles 4.4}: A.Kwiatkowski, H.Spiesberger and H.J.M\"ohring,
Proc.~1991 Workshop on Physics at HERA, ed. W.Buchm\"uller and G.Ingelman
(DESY, Hamburg, 1992), Vol.~3, p.1294.

\bibitem{django} {\sc django 6.1}: G.Schuler and H.Spiesberger,
Proc.~1991 Workshop on Physics at HERA, ed. W.Buchm\"uller and G.Ingelman
(DESY, Hamburg, 1992), Vol.~3, p.1419.

\bibitem{mrsa} A.D.Martin, R.G.Roberts, W.J.Stirling, Phys.~Rev.~{\bf D50}(1994)6734.

\bibitem{ariadne} {\sc ariadne 4.06}: L.L\"onnblad, LU TP--89--10;\\
                  L.L\"onnblad, Comp.~Phys.~Comm.~{\bf 71}(1992)15.

\bibitem{pythia} {\sc pythia 5.7} and {\sc jetset 7.4}: T.Sj\"ostrand, CERN--TH 7112--93 (1994);\\
                 T.Sj\"ostrand, LU--TP--95--20 (1995).

\bibitem{herwig} {\sc herwig 5.8}: B.R.Webber,
Proc.~1991 Workshop on Physics at HERA, ed. W.Buchm\"uller and G.Ingelman
(DESY, Hamburg, 1992), Vol.~3, p.1363.

\bibitem{aroma} {\sc aroma 2.1}: G.Ingelman and G.Schuler,
Proc.~1991 Workshop on Physics at HERA, ed. W.Buchm\"uller and G.Ingelman
(DESY, Hamburg, 1992), Vol.~3, p.1346.

\bibitem{pompyt} {\sc pompyt 1.0}: P.Bruni and G.Ingelman, Proc.~Europhysics Conference on HEP, 
Marseilles, France (1993), p.595.

\bibitem{Schlein} G.Ingelman and P.Schlein, Phys.~Lett.~{\bf B152}(1985)256.

\bibitem{geant} R.Brun et al., CERN DD/EE--84--1 (1987).

\bibitem{JB} F.Jacquet and A.Blondel, Proceedings of the study for an
$ep$ facility for Europe, DESY 79--48 (1979), p.391.

\bibitem{sinistra} H.Abramowicz, A.Caldwell, R.Sinkus, Nucl.~Inst.~Meth.~{\bf A365}(1995)508.

\bibitem{DA} S.Bentvelsen, J.Engelen, P.Kooijman, 
Proc.~1991 Workshop on Physics at HERA, ed. W.Buchm\"uller and G.Ingelman
(DESY, Hamburg, 1992), Vol.~1, p.23.

\bibitem{Julio} G.D'Agostini, DESY~94--099 (1994).

\bibitem{MRSD0} A.D.Martin, R.G.Roberts, W.J.Stirling, Phys.~Rev.~{\bf D47}(1993)867.

\bibitem{MRSDmnsp} A.D.Martin, R.G.Roberts, W.J.Stirling, Phys.~Lett.~{\bf306B}(1993)147;\\
  erratum in Phys.~Lett.~{\bf309B}(1993)492.

\bibitem{GRV} M.Gl\"uck, E.Reya, A.Vogt, Z.~Phys.~{\bf C53}(1992)127.

\bibitem{CTEQ} CTEQ Collaboration, J.Botts et al., Phys.~Lett.~{\bf B304}(1993)159.

\bibitem{BEBCdoveru} BEBC WA21 Collaboration, G.T.Jones et al., Z.~Phys.~{\bf C62}(1994)601.

\bibitem{PDG95} L.Montanet et al., Phys.~Rev.~{\bf D50}(1994)1173.

\bibitem{snowmass} UA1 Collaboration, G.Arnison et al., Phys.~Lett.~{\bf B123}(1983)115;\\
  J.Huth et al., Proc.\ 1990 DPF Summer Study on High--Energy Physics, Snowmass, Colorado,
  ed. E.L.Berger (World Scientific, Singapore, 1990, p.134).

\bibitem{ZEUSjets} ZEUS Collaboration, M.Derrick et al., Z.~Phys.~{\bf C67}(1995)81.

\bibitem{DISVecMes} ZEUS Collaboration, M.Derrick et al., Phys.~Lett.~{\bf B356}(1995)601;\\
                    H1   Collaboration, S.Aid et al., DESY 96--023.

\bibitem{SoftCol} A.Edin, G.Ingelman, J.Rathsman, DESY 95-145 (1995).

\end{thebibliography}
\end{document}